\documentclass[superscriptaddress,nofootinbib,twocolumn]{revtex4}
\usepackage{amsmath,lscape,epsfig}
\usepackage{amsfonts}
\usepackage{amsmath}
\usepackage{amssymb}
\usepackage{slashed}
\usepackage{color,soul}
\usepackage[utf8]{inputenc}
\usepackage{graphicx}
\usepackage{epstopdf}
\usepackage{ulem}

\DeclareMathOperator{\Tr}{Tr}

\def\beq{\begin{equation}}
\def\eeq{\end{equation}}
\def\beqa{\begin{eqnarray}}
\def\eeqa{\end{eqnarray}}
\def\ban{\begin{eqnarray*}}
\def\ean{\end{eqnarray*}}
\def\bi{\begin{itemize}}
\def\ei{\end{itemize}}

\begin{document}

\title{Deconfinement and chiral phase transitions in quark matter with a strong electric field}

\author{William R. Tavares} \email{william.tavares@posgrad.ufsc.br}
\affiliation{Departamento de F\'{\i}sica, Universidade Federal de Santa
  Catarina, 88040-900 Florian\'{o}polis, Santa Catarina, Brazil}  

\author{Ricardo L. S. Farias}
\email{ricardo.farias@ufsm.br}
\affiliation{Departamento de F\'{\i}sica, Universidade Federal de Santa Maria,
97105-900 Santa Maria, RS, Brazil}

\author{Sidney S. Avancini} \email{sidney.avancini@ufsc.br}
\affiliation{Departamento de F\'{\i}sica, Universidade Federal de Santa
  Catarina, 88040-900 Florian\'{o}polis, Santa Catarina, Brazil}

\begin{abstract}

The deconfinement and chiral phase transitions are studied in the context of the electrized quark matter at finite temperature in 
the two-flavor Polyakov--Nambu--Jona-Lasinio model. Using the mean field approximation and an electric field independent regularization
we show that the effect of temperature and/or electric fields is to partially restore the chiral symmetry. The deconfinement 
phase transition is slightly affected by the magnitude of the electric field. 
To this end we show how the effective quark masses
and the expectation value of the Polyakov loop are affected by the electric fields at finite temperatures. As a very interesting 
result, the pseudocritial temperatures for chiral symmetry restoration and deconfinement decrease as we increase 
the magnitude of the electric fields, however,  both start to increase  after some critical value of the electric field. 
 \end{abstract}

\maketitle

\section{Introduction}
\label{intro}
Recent numerical simulations provide possibilities for strong electromagnetic fields to be present in noncentral
heavy ion collisions (HIC)~\cite{skokov,deng,zhang,deng2}. These indications suggest even more properties to be explored
 in the strongly interacting quark matter, besides the usual strong magnetic fields~\cite{fukushima2}, that are supposed to be created in noncentral HIC or in magnetars~\cite{duncan,kouve}.
The event-by-event fluctuations of the proton positions in the colliding nuclei in Au+Au heavy-ion collisions at
$\sqrt{s}=200$ GeV in RHIC@BNL and Pb+Pb at $\sqrt{s}=2.76$ TeV in ALICE@LHC, indicate the creation of strong electric fields that 
can be of the same order of magnitude of the already predicted magnetic fields~\cite{skokov,deng,zhang,zhang2}. In asymmetric
Cu+Au collisions~\cite{hirono,deng2,Voronyuk} 
it is expected that a strong electric field is generated in the overlapping region. This happens because there is a different number
of electric charges in each nuclei and it is argued that this is a fundamental property due to the charge dipole formed in the early 
stage of the collision. Also, different projectile-target combinations are studied from symmetric to asymmetric collisions systems, 
showing that the electric field is more significant in the former~\cite{cheng2}. The time evolution for electric and magnetic fields in HIC can be estimated and the prediction depends on the conductivity of the medium~\cite{sheng}. In a complementary way, 
the study 
of strong electric fields can be very useful when searching for the chiral magnetic effect (CME)~\cite{fukushima2},  
given the speculated possibility of reversing the sign of some experimental observables related to the CME ~\cite{deng2,review} 
if the lifetime of such fields is long enough. Such 
scenarios could give an opportunity to explore anomalous transport properties such as chiral electric separation~\cite{xu1,xu2,shipu,shipu2}, which 
is a generation of an axial current in a system with both vector and axial densities~\cite{review2}. 

It is  natural in this scenario to ask how some properties of quantum chromodynamics(QCD) under 
strong conditions (i.e.., high temperatures and densities) can be affected by such fields. In the low energy limit 
we should use effective theories or lattice QCD (LQCD) techniques, once we are dealing with the nonperturbative behavior of QCD.
One of the main aspects that should be explored is the
chiral symmetry restoration of QCD, where several studies have explored a series of interesting phenomena, 
like the magnetic catalysis~\cite{shovkovy,nosso1,avancini2}, the chiral magnetic effect~\cite{fukushima2} and the 
inverse magnetic catalysis~\cite{Farias:2014eca, Farias:2016gmy,Avancini:2016fgq,Avancini:2018svs,Ayala_LSM1,Ayala_LSM2}, predicted by 
lattice QCD results~\cite{bali,endrodiIMC}, all of them studied in the context of a pure magnetic field. 
A natural extension of these works can be done by exploring the role of magnetic fields in the deconfinement 
transition as well~\cite{gatto1,gatto2,gatto3,kashiwa,endrodi,fukushima}.  The effective theories or LQCD investigations in general are
dealing with the chiral condensate, which is an approximate order parameter for the 
chiral phase transition~\cite{fukushima}. On the other hand, the deconfinement in a gauge theory
is associated with the spontaneous symmetry breaking of the center symmetry, 
in which the approximate order parameter is
the Polyakov loop~\cite{mclerran,cheng}. The predictions 
of some effective models for the physics associated with the Polyakov loop have been widely studied in the literature~\cite{andersen0,ratti01,ratti0,fukushima03,ratti}.

The main goal of our work is to include for the first time the effects of a pure electric field on the Polyakov
extended SU(2) Nambu--Jona-Lasinio model (PNJL)~\cite{nambu}, to
study how the deconfinement phase transition temperature $T^l_{pc}$ 
is affected by these fields, and also to explore its connection 
with the chiral pseudocritial temperature $T^{\chi}_{pc}$ . 
To this end, we will implement the Schwinger proper-time quark propagators
in a constant electric field, which have been explored in previous works~\cite{elec1,Cao,ruggieri1,ruggieri2} 
 related to the study of the gap equation and
thermodynamical properties of the model. We also implement in 
the Schwinger proper-time formalism the Polyakov loop, which should be useful for future
evaluations even at $eE=0$. As the Nambu--Jona-Lasinio SU(2) model (NJL) in 3+1 D is nonrenormalizable, 
the regularization procedure is the same as that presented in~\cite{elec1}, where the pure-electric field contribution 
 was analytically solved and the finite thermoelectric part
 was numerically evaluated.
 Finally, the Schwinger pair production~\cite{Schwinger,heisenberg} will
be presented. 
Previous studies have shown~\cite{elec1,Cao} that in the present approximation this quantity is mainly 
determined by the 
effective quark masses, which in turn incorporates the effects  of temperature 
and electric field. Thus, there will be quantitative differences between the results for the pair production rate when calculated
in the PNJL or NJL model.

The work is organized as follows. In Sec.~\ref{formalism} we introduce the standard formalism for the two-flavor NJL and PNJL 
models at finite temperature including the effects due to the strong electric field. 
In Sec.~\ref{reg} we present the
regularization scheme adopted in this work.  In Sec.~\ref{numericalres} we present our numerical results. Finally, in Sec.~\ref{conclusion}
we discuss our results and conclusions. We leave for the Appendix the explicit calculations of 
the trace of the Polyakov loop.

\section{General formalism}
\label{formalism}

In the presence of an electromagnetic field the two-flavor NJL model Lagrangian can be written as

\begin{eqnarray}
\mathcal{L}&=&\overline{\psi}\left(i \slashed D - \tilde{m}\right)\psi
+G\left[(\overline{\psi}\psi)^{2}+(\overline{\psi}i\gamma_{5}\vec{\tau}\psi)^{2}\right]\nonumber\\
&&-\frac{1}{4}F^{\mu\nu}F_{\mu\nu}~,\label{su3}
\end{eqnarray}

\noindent where $A^\mu$, $F^{\mu\nu} = \partial^\mu A^\nu - \partial^\nu A^\mu$ 
are the electromagnetic gauge potential and tensor fields, respectively; 
$G$ is the coupling 
constant; and
$D^\mu =(i\partial^{\mu}-QA^{\mu})$ is the covariant derivative.

Also, $\vec{\tau}$ are the isospin Pauli 
matrices; Q is the diagonal quark charge 
 matrix, 
Q=diag($q_u$= $2 e/3$, $q_d$=-$e/3$);  
 $\psi=(\psi_u,\psi_d)^T$ is the two-flavor quark fermion field; and $\tilde{m}=$diag($m_u$,$m_d$) represents 
 the current quark mass matrix. Here, we adopt the isospin approximation, i.e., $m_u=m_d=m$. 
 We choose $A_{\mu}=-\delta_{\mu 0}x_{3}E$ in order to obtain a
 constant electric field in the z-direction.

In the mean field approximation the Lagrangian density reads
\begin{equation}
 \mathcal{L}=\overline{\psi}\left(i\slashed D-M\right)\psi+G \left \langle \overline{\psi}\psi \right \rangle^{2}-
 \frac{1}{4}F^{\mu\nu}F_{\mu\nu}~,
\end{equation}
\noindent where the constituent quark mass is defined by the following expression:

\begin{eqnarray}
M=m-2G\sum_{f=u,d} \left(\phi_f^{\mathcal{E}}+\phi_f^{\mathcal{E},T}\right), \label{gap2} 
\end{eqnarray}

\noindent where we have both  a pure electric field, $\phi_f^{\mathcal{E}}$, and a thermoelectric field, $\phi_f^{\mathcal{E},T}$, contribution.
 We will use the following definition of the quark condensate:

\begin{eqnarray}
\phi_f=\left \langle \overline{\psi}_f\psi_f \right \rangle =-\int\frac{d^4p}{(2\pi)^4} 
\Tr\left[iS_f(p)\right],\label{cond1}
\end{eqnarray}
where $f=u,d$ stands for the quark flavors.
Following the procedure described in~\cite{elec1}, the pure electric condensate contribution $\phi_f^{\mathcal{E}}$ can be calculated using
the full Schwinger proper-time quark propagator~\cite{Schwinger} in a constant electric field, resulting in the expression:

\begin{equation}
 \phi_f^{\mathcal{E}}=-\frac{MN_c}{4\pi^2}\mathcal{E}_f\int_0^{\infty}ds\frac{e^{-sM^2_f}}{s}
 \cot(\mathcal{E}_f s) ~, \label{phisu3}
\end{equation}
where $\mathcal{E}_f=|q_f|E$ and $N_c=3$ is the number of colors.

As already derived in~\cite{elec1,Cao}, the thermoelectric contribution can be written as:

\begin{eqnarray}
 &&\phi_f^{\mathcal{E},T}=- \frac{MN_c}{2\pi^2}\sum_{n=1}^{\infty}(-1)^n\mathcal{E}_f\int_0^{\infty}ds\frac{e^{-sM^2_f}}{s}\cot(\mathcal{E}_fs)\nonumber\\
 &&\times e^{-\frac{\mathcal{E}_fn^2}{4|\tan(\mathcal{E}_fs)|T^2}}. \label{phi001}
\end{eqnarray}

The thermodynamical potential has been derived in~\cite{Cao,elec1} 
 simply by integrating Eq.~(\ref{gap2}) with respect to  the
effective quark mass $M$, yielding

\begin{eqnarray}\label{pot01}
\Omega=\frac{(M-m)^2}{4G}-\sum_{f=u,d}\left(\theta^{\mathcal{E}}_f+
\theta_f^{\mathcal{E},T} \right),
\end{eqnarray}
\noindent where we have defined $\theta_f^\mathcal{E}$ and $\theta_f^{\mathcal{E},T}$ as

\begin{eqnarray}
 \theta_f^\mathcal{E}&=&-\frac{N_c}{8\pi^2}\int_0^{\infty}ds\frac{e^{-sM^2}}{s^2}\mathcal{E}_f\cot(\mathcal{E}_fs),\label{thetaE}\\
 \theta_f^{\mathcal{E},T}&=&-\frac{N_c}{4\pi^2}\sum_{n=1}^\infty(-1)^n\int_0^{\infty}ds\frac{e^{-sM^2}}{s^2}\mathcal{E}_f\cot(\mathcal{E}_fs)\nonumber\\
 &&\times e^{-\frac{\mathcal{E}_fn^2}{4|\tan(\mathcal{E}_fs)|T^2}}.\label{thetaET0}
\end{eqnarray}

The Schwinger pair production rate is given by $\Gamma=-2\Im\left(\Omega\right)$~\cite{Cao,Schwinger}, where 
$\Im\left(\Omega\right)$ corresponds to the imaginary part of the effective potential. Explicitly, one obtains for  $\Gamma$

 \begin{equation}
  \Gamma(M,\mathcal{E},T)=\frac{N_c}{4\pi}\sum_{f}\mathcal{E}_f^2\sum_{k=1}^{\infty}\frac{e^{-\frac{M^2\pi k}{\mathcal{E}_f} }}{(k\pi)^2},\label{decay}
 \end{equation}

\noindent where we need to perform the summation over the flavor indices $f=u,d$, and
as we will see later, the entire Schwinger 
pair production dependence on the external conditions is contained  only in the effective 
mass $M\equiv M(\mathcal{E},T)$.

%
%

\subsection{The electrized SU(2) PNJL}
The extended version of the two-flavor NJL model Lagrangian in the 
presence of a electromagnetic field coupled with the Polyakov loop is given by

\begin{eqnarray}
\mathcal{L}=&&\overline{\psi}\left(i \slashed D - \tilde{m}\right)\psi
+G\left[(\overline{\psi}\psi)^{2}+(\overline{\psi}i\gamma_{5}\vec{\tau}\psi)^{2}\right]\nonumber\\
&&-\frac{1}{4}F^{\mu\nu}F_{\mu\nu} -\mathcal{U}(l,\overline{l},T) ~,\label{su3pnjl}
\end{eqnarray}

\noindent now the covariant derivative is given by $D^\mu =(i\partial^{\mu}-QA^{\mu}-i\mathcal{A}^{\mu})$, 
 where $\mathcal{A}^{\mu}=\delta^{\mu}_0\mathcal{A}^0$ is the Polyakov gauge,
the strong coupling constant $g$ is absorbed in the definition $\mathcal{A}^{\mu}(x)=g\frac{\lambda_a}{2}\mathcal{A}_a^{\mu}(x)$,
$\lambda_a$ are the Gell-Mann matrices and $\mathcal{A}_a^{\mu}(x)$ is the SU(3) gauge field. 

For the pure gauge sector, let us define the Polyakov line as
\begin{eqnarray}
 L(x)=\mathcal{P}\exp\left[i\int_0^{\beta}d\tau\mathcal{A}_4(\tau,\overrightarrow {x}) \right],
\end{eqnarray}
\noindent where $\mathcal{P}$ is a path ordering and $\beta=\frac{1}{T}$. Also, $\mathcal{A}_4=i\mathcal{A}_0$ is the temporal 
component of the Euclidean gauge field 
$(\mathcal{A}_4,\overrightarrow{\mathcal{A}})$.

The effective potential $\mathcal{U}(l,\overline{l},T)$ for the Polyakov fields is parametrized in 
order to reproduce lattice results 
in the mean field 
approximation~\cite{ratti01,ratti0}. We adopt the Ansatz~\cite{ratti0}, 

\begin{eqnarray}
\frac{\mathcal{U}(l,\overline{l},T)}{T^4}=-\frac{b_2(T)l\overline{l}}{2} -\frac{b_3(l^3+\overline{l}^3)}{6}+\frac{b_4(l\overline{l})^2}{4},\label{potPNJL}
\end{eqnarray}
\noindent where $b_2(T)$ is given by

\begin{eqnarray}
 b_2(T)=a_0+a_1\left(\frac{T_0}{T}\right)+a_2\left(\frac{T_0}{T}\right)^2+a_3\left(\frac{T_0}{T}\right)^3.
\end{eqnarray}

 The parameters of the potential $\mathcal{U}(l,\overline{l},T)$ 
will be given Section \ref{numericalres}. 
We should also mention that the thermal expectation value of the Polyakov loop is given by~\cite{fukushima03}

\begin{eqnarray}
\l \equiv \frac{1}{N_c} \left \langle Tr_c L(x) \right \rangle, \quad \overline{l} 
\equiv \frac{1}{N_c} \left \langle Tr_c L^\dagger(x) \right \rangle .
\end{eqnarray}

 As we will see,
 our most interesting results occur temperatures around the pseudocritial temperatures for 
 deconfinment $T^{l}_{pc}$ and for chiral symmetry restoration $T^{\chi}_{pc}$. We verify that the different 
 {\it Ansätze} for $U(l,\bar{l},T)$~\cite{fukuansatz}  almost agree in this region. 
 Therefore, our results remain essentially the same if we change the form of the  effective potential for the Polyakov loop $U(l,\bar{l},T)$.
 
Once we are working with the background field $\mathcal{A}_4$, we can 
 obtain the condensate as a straightforward generalization of the expression
given in Eq.~(\ref{cond1}) for zero 
temperature and density   using
 the following symbolic replacements~\cite{ratti}: 

\begin{eqnarray}
 i\int\frac{d^4p}{(2\pi)^4}&\rightarrow&-T\frac{1}{N_c}\Tr_c\sum_{n=-\infty}^{\infty}\int\frac{d^3p}{(2\pi)^3},\label{intrep}\\
 (p_0,\overrightarrow{p})&\rightarrow& (i\omega_n+\mu-i\mathcal{A}_4,\overrightarrow{p}),\label{repl}
\end{eqnarray}

\noindent where $w_n=(2n+1)\pi T$ is the Matsubara frequency. 
 In this work we consider only the zero baryon chemical potential case, hence $l$=$\overline{l}$.
  Once the traced Polyakov loop is given by $l=\frac{1}{N_c}Tr\exp(i\frac{\mathcal{A}_4}{T})$,
we can write the background field in the Polyakov gauge\cite{andersen0, lowe} as $\mathcal{A}_4=g\mathcal{A}_4^{(3)}\frac{\lambda_3}{2}+g\mathcal{A}_4^{(8)}\frac{\lambda_8}{2}$, and it is straightforward to see 
that $\mathcal{A}_4^{(8)}=0$ at $\mu=0$.
Therefore, we implement the Polyakov loop in the condensate $\phi^{\mathcal{E},T,l}$ at $\mu=0$ by 
using the prescriptions given by 
Eqs. (\ref{intrep}) and (\ref{repl}). After calculating  the trace in the color space, one obtains

\begin{eqnarray}
 &&\phi_f^{\mathcal{E},T,l}=- \frac{M}{2\pi^2}\sum_{n=1}^{\infty}(-1)^n\mathcal{E}_f\int_0^{\infty}ds
 \frac{e^{-sM^2}}{s}\cot(\mathcal{E}_fs)\nonumber\\
 &&\times e^{-\frac{\mathcal{E}_fn^2}{4|\tan(\mathcal{E}_fs)|T^2}}\left\{1+2\cos\left[n\cos^{-1}\left(\frac{3l-1}{2}\right)\right] \right\}.\label{phi00}\nonumber\\
\end{eqnarray}

By making use of this expression in Eq.(\ref{gap2})  
the SU(2) PNJL gap equation reads

\begin{equation}
 \frac{M-m}{2G}=-\sum_{f=u,d}\left(\phi^{\mathcal{E}}_f+
 \phi_f^{\mathcal{E},T,l}\right).\label{gap2PNJL}
\end{equation}

 Finally, the thermodynamical or effective potential  is obtained following a 
standard procedure~\cite{ratti}
using the prescriptions (\ref{intrep}) and (\ref{repl}), yielding

\begin{eqnarray}
\Omega=\mathcal{U}(l,T)+\frac{(M-m)^2}{4G}-\sum_{f=u,d}\left(\theta^{\mathcal{E}}_f+
\theta_f^{\mathcal{E},T,l} \right).\label{pot01PNJL}
\end{eqnarray}

It is straightforward to show that : 
\begin{eqnarray}
 &&\theta_f^\mathcal{E}=-\frac{N_c}{8\pi^2}\int_0^{\infty}ds\frac{e^{-sM^2}}{s^2}\mathcal{E}_f\cot(\mathcal{E}_fs),
 \label{thetaE}\\
 &&\theta_f^{\mathcal{E},T,l}=-\frac{1}{4\pi^2}\sum_{n=1}^\infty(-1)^n\int_0^{\infty}ds\frac{e^{-sM^2}}{s^2}
 \mathcal{E}_f\cot(\mathcal{E}_fs)\nonumber\\
 &&\times e^{-\frac{\mathcal{E}_fn^2}{4|\tan(\mathcal{E}_fs)|T^2}}\left\{1+2\cos\left[n\cos^{-1}
 \left(\frac{3l-1}{2}\right)\right] \right\}.\nonumber \\
 \label{thetaET} \, .
\end{eqnarray}

The effective quark masses and the expectation value of the Polyakov loop are obtained by minimizing the thermodynamical potential, i e.,
calculating 
 $\frac{\partial \Omega}{\partial M}=0$ and $\frac{\partial \Omega}{\partial l}=0$. 
 The derivative $\frac{\partial \Omega}{\partial M}=0$  was
already obtained in Eq.(\ref{gap2PNJL}) and $\frac{\partial \Omega}{\partial l}=0$ is given by

\begin{eqnarray}
&&0=\frac{T^4}{2}\left[-b_2(T)l-b_3l^2+b_4l^3\right]+\nonumber\\ 
&&\frac{1}{4\pi^2}\sum_{f=u,d}\int_{0}^{\infty}\frac{ds}{s^2}e^{-sM^2}\mathcal{E}_f\cot(\mathcal{E}_fs)\sum_{n=0}^{\infty}(-1)^n \nonumber\\
&&\times e^{-\frac{\mathcal{E}_fn^2}{4|\tan(\mathcal{E}_fs)|T^2}}\left\{\frac{\sqrt{3}n\sin [n\cos^{-1}\left(\frac{3l-1}{2}\right)]}{\sqrt{-3l^2+2l+1}} \right\}.\nonumber \\
\label{termgap}
\end{eqnarray}
\section{Regularization}
\label{reg}
In this work we use the vacuum subtraction scheme~\cite{Cao,ruggieri1,ruggieri2,elec1}. We define for 
the vacuum-subtracted condensate the following quantity:
\begin{eqnarray}
\overline{\phi}_f^{\mathcal{E}}=-\frac{MN_c}{4\pi^2}\int_0^{\infty}ds
\frac{e^{-sM^2}}{s^2}\left[\mathcal{E}_fs\cot(\mathcal{E}_fs)-1\right] \, . \label{condE4}
\end{eqnarray}
The gap equation, Eq.(\ref{gap2}) in the SU(2) NJL model and Eq.(\ref{gap2PNJL}) in the SU(2) PNJL, 
should be calculated using the 
following regularized condensate $\phi_f^{\mathcal{E}}$:
\begin{equation}
 \phi_f^{\mathcal{E}}=\overline{\phi}_f^{\mathcal{E}}+\phi_f^{vac} \, ,
\end{equation}

\noindent where we have adopted the 3D cutoff scheme~\cite{kleva,buballa}  
 to regularize the infinite field independent vacuum contribution

\begin{equation}
\phi_f^{vac}=-\frac{MN_c}{2\pi^2}\left[\Lambda \mathcal{E}_{\Lambda} -
      M^2\ln\left( \frac{\mathcal{E}_{\Lambda}+\Lambda}{M} \right)\right],
\end{equation}
where $\mathcal{E}_{\Lambda}=\sqrt{\Lambda^2+M^2_f}$.
In the same way, for the effective potential in the SU(2) NJL model, Eq.(\ref{pot01}), and 
Eq.(\ref{pot01PNJL}) in the SU(2) PNJL model, 
we should use the regularized $\theta_f^{\mathcal{E}}$ given by
\begin{equation}
 \theta_f^{\mathcal{E}}=\overline{\theta}_f^{\mathcal{E}} +\theta_f^{vac} \, ,
\end{equation}
\noindent where the finite field dependent term is
{\small \begin{eqnarray}
  \overline{\theta}_f^{\mathcal{E}}=-\frac{N_c}{8\pi^2}\int_0^{\infty}ds\frac{e^{-sM^2}}{s^3} \label{thetaE4}
  \left[\mathcal{E}_fs\cot(\mathcal{E}_fs)-1+\frac{(\mathcal{E}_f s)^2}{3}\right],
\end{eqnarray}} 
\noindent and the infinite vacuum contribution $\theta_f^{vac}$ is given by
\begin{equation}
\theta_f^{vac}=\frac{N_c}{8\pi^2}\int_{0}^{\infty} ds \frac{e^{-sM^2}}{s^3}.\label{3d}
\end{equation}

\noindent The regularized $\theta_f^{vac}$ is given in the 3D cutoff scheme by  
\begin{eqnarray}
\theta^{vac}_f&=&-\frac{N_c}{8\pi^2}\left[M^4\ln\left(\frac{\Lambda+\sqrt{\Lambda^2+M^2}}{M}\right)\right.
\nonumber\\
&&-\left. \Lambda\sqrt{\Lambda^2+M^2}\left(M^2+2\Lambda^2\right) \right].\label{omegavac}
\end{eqnarray}

It's important to note that the poles associated with the imaginary part of the
effective potential, can be associated to 
the zeros of the $\sin(\mathcal{E}_fs)$ which appear in the denominator of
both our gap equation and the effective potential  
when $\mathcal{E}_fs=n\pi$ for $n=1,2,3,...$ . 
 We interpret these integrals as the Cauchy principal value~\cite{elec1,ruggieri1} and that the 
 true ground state of the theory
is given by the real part of the thermodynamical potential.

Using the techniques adopted in~\cite{elec1}, we can use the principal value (or the real part) of 
$\overline{\theta}_f^{\mathcal{E}}$, given by 

\begin{eqnarray}
\Re \left( \overline{\theta}_f^{\mathcal{E}} \right) &=&-\frac{N_c}{2\pi^2}(\mathcal{E}_f)^2 
\left \{ \zeta'(-1) + \frac{\pi}{4}y_f + \frac{y_f^2}{2} \left( \gamma_E-\frac{3}{2} \right.\right.\nonumber \\
&&+\left.\left.\ln y_f \right) - \frac{1}{12} \left(1+ \ln y_f \right) +\sum_{k=1}^{\infty}k\left[\frac{y_f}{k} \right.\right. \nonumber\\ 
&&\left.\left.\times\tan^{-1}\left(\frac{y_f}{k}\right) - \frac{1}{2}\ln\left(1+\left(\frac{y_f}{k}\right)^2\right)\right.\right.\nonumber\\
&&-\left.\left.\frac{1}{2}\left(\frac{y_f}{k}\right)^2\right] \right \}, 
\label{potef1}
\end{eqnarray} 
\noindent where $y_f=M^2/(2\mathcal{E}_f)$. In the same manner, we obtain 
for the principal value
of the vacuum-subtracted condensate $\overline{\phi}_f^{\mathcal{E}}$
\begin{eqnarray}
\Re \left( \overline{\phi}_f^{\mathcal{E}} \right) &=&-\frac{MN_c}{4\pi^2}\int_0^{\infty}ds
\frac{e^{-sM^2}}{s^2}\left[\mathcal{E}_fs\cot\left(\mathcal{E}_fs\right)-1\right]   \nonumber \\ 
&&=\frac{MN_c}{2\pi^2}\mathcal{E}_f \left[ \frac{\pi}{4}+y_{f}(\gamma_E-1+\ln y_{f})+\right. \nonumber \\
&&+\left. \sum_{k=1}^{\infty}\left(\tan^{-1}\frac{y_{f}}{k}-\frac{y_{f}}{k} \right)\right]. \label{reg1}
\end{eqnarray}

As discussed in~\cite{elec1}, the quantities $\phi_f^{\mathcal{E},T}$ and $\theta_f^{\mathcal{E},T}$ depend on 
the temperature and following the procedure adopted in Ref.~\cite{ayala} we set the lower 
limits of the integration to zero, since theses integrals are ultraviolet finite.

%
%
%
%
%
\section{Numerical Results}
\label{numericalres}
In this section we show our numerical results. The parameter set for the SU(2) NJL model is~\cite{buballa} $\Lambda=587.9$ MeV, $G\Lambda^2=2.44$, and
$m=5.6$ MeV. The parameters associated with the pure gauge sector are the following:  $a_0=6.75$, $a_1=-1.95$, $a_2=2.625$, $a_3=-7.44$, $b_3=0.75$, and $b_4=7.5$. In the pure gauge sector, the transition temperature is given by $T_0=270$ MeV~\cite{ratti0}. 
A lower value of $T_0$ is usually necessary in order to include the effects of
$N_f=2$ in the theory. Following the procedure adopted in Refs.~\cite{paw,endrodi}, we use $T_0=208$ MeV.
\begin{center}
\begin{figure}[!htb]
\includegraphics[width=7cm]{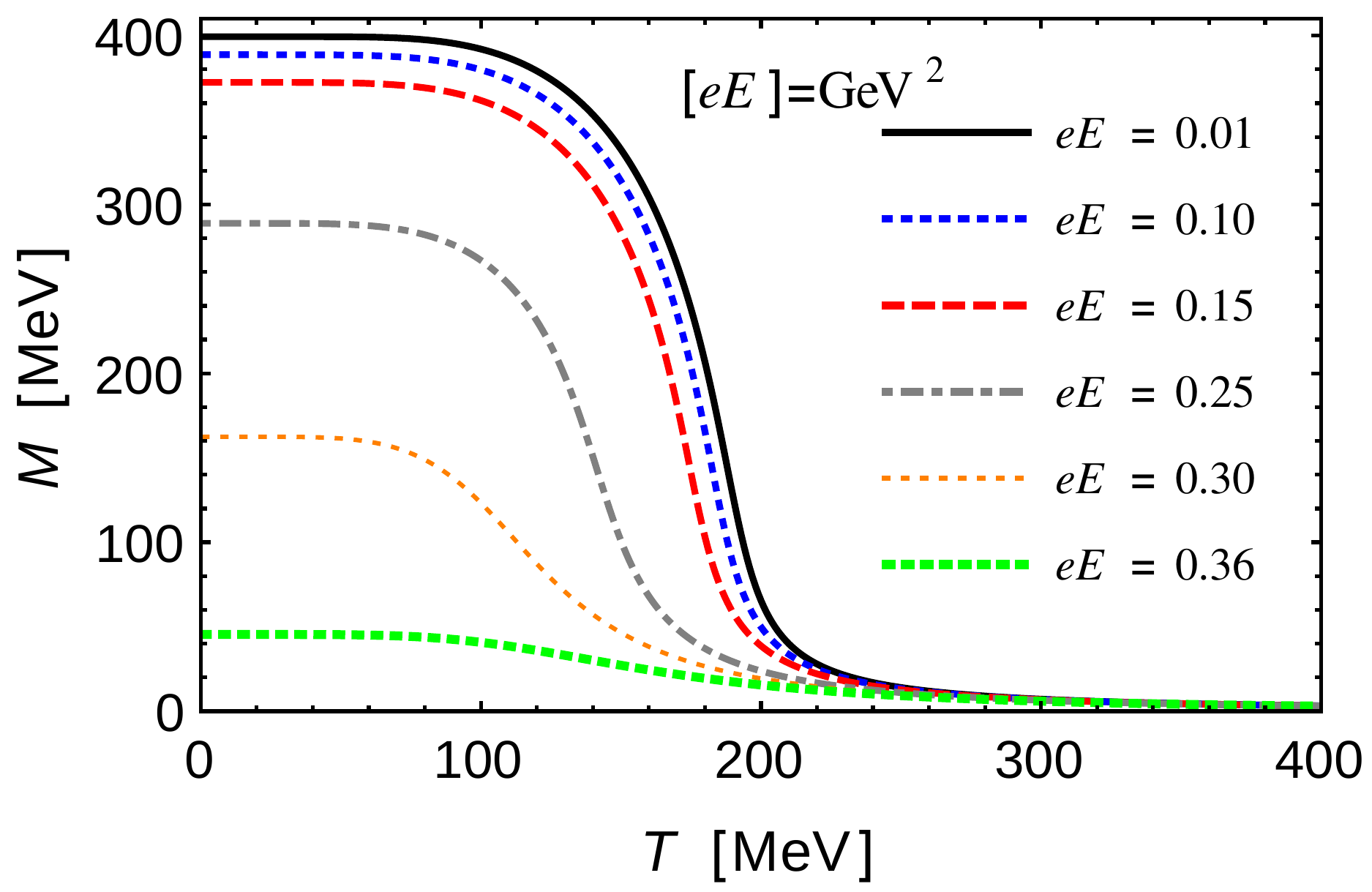}
\caption{Constituent quark mass $M$ as a function of the temperature for different values of $eE$ in 
the SU(2) NJL model.}
\label{fig1}
\end{figure}
\end{center}

\begin{center}
\begin{figure}[!htb]
\includegraphics[width=7cm]{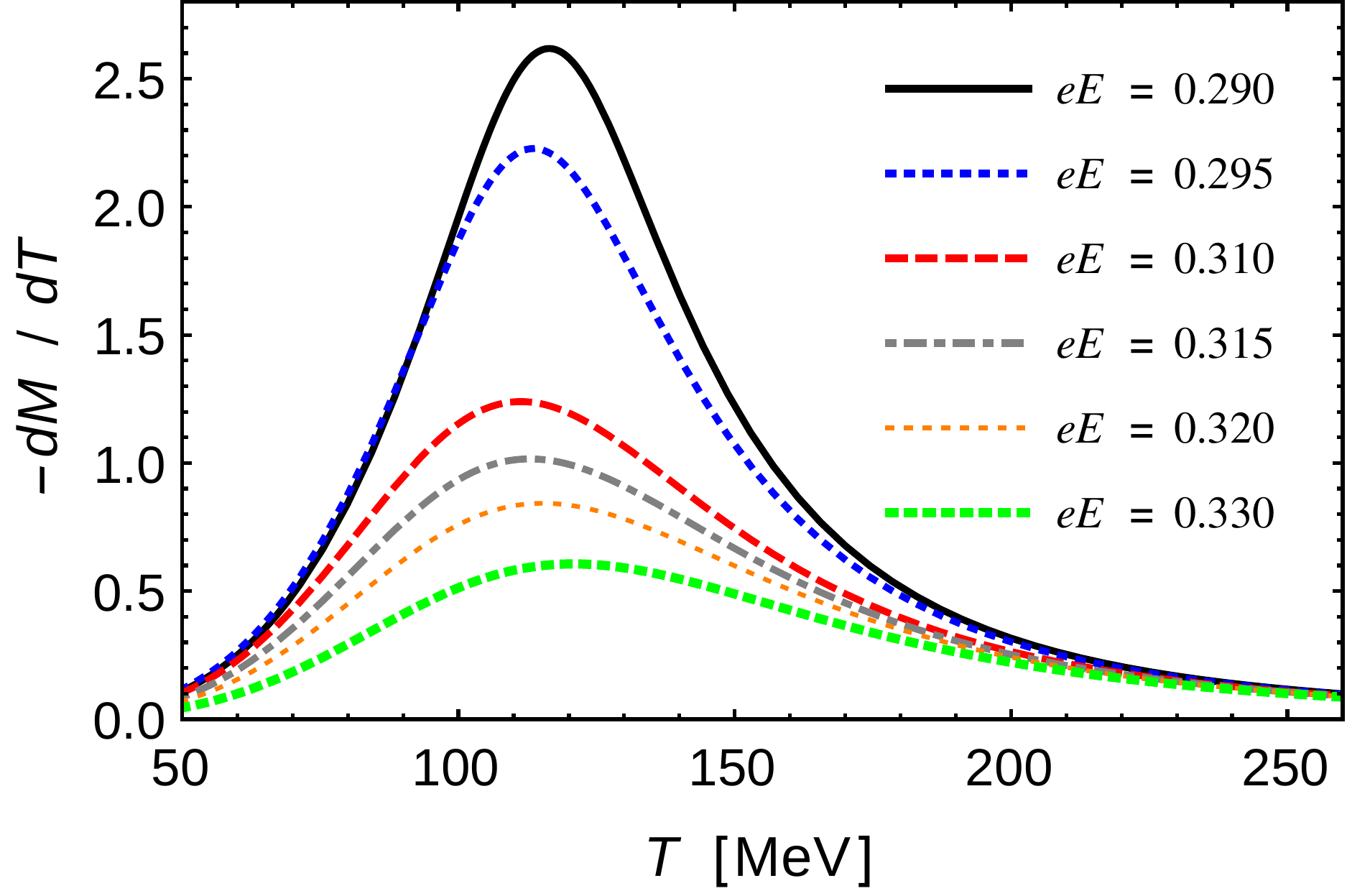}
\caption{Thermal susceptibility $- \frac{dM}{dT}$ as a function of the temperature for different
values of $eE$ in the SU(2) NJL model.}
\label{fig2}
\end{figure}
\end{center}

\begin{center}
\begin{figure}[!htb]
\includegraphics[width=7cm]{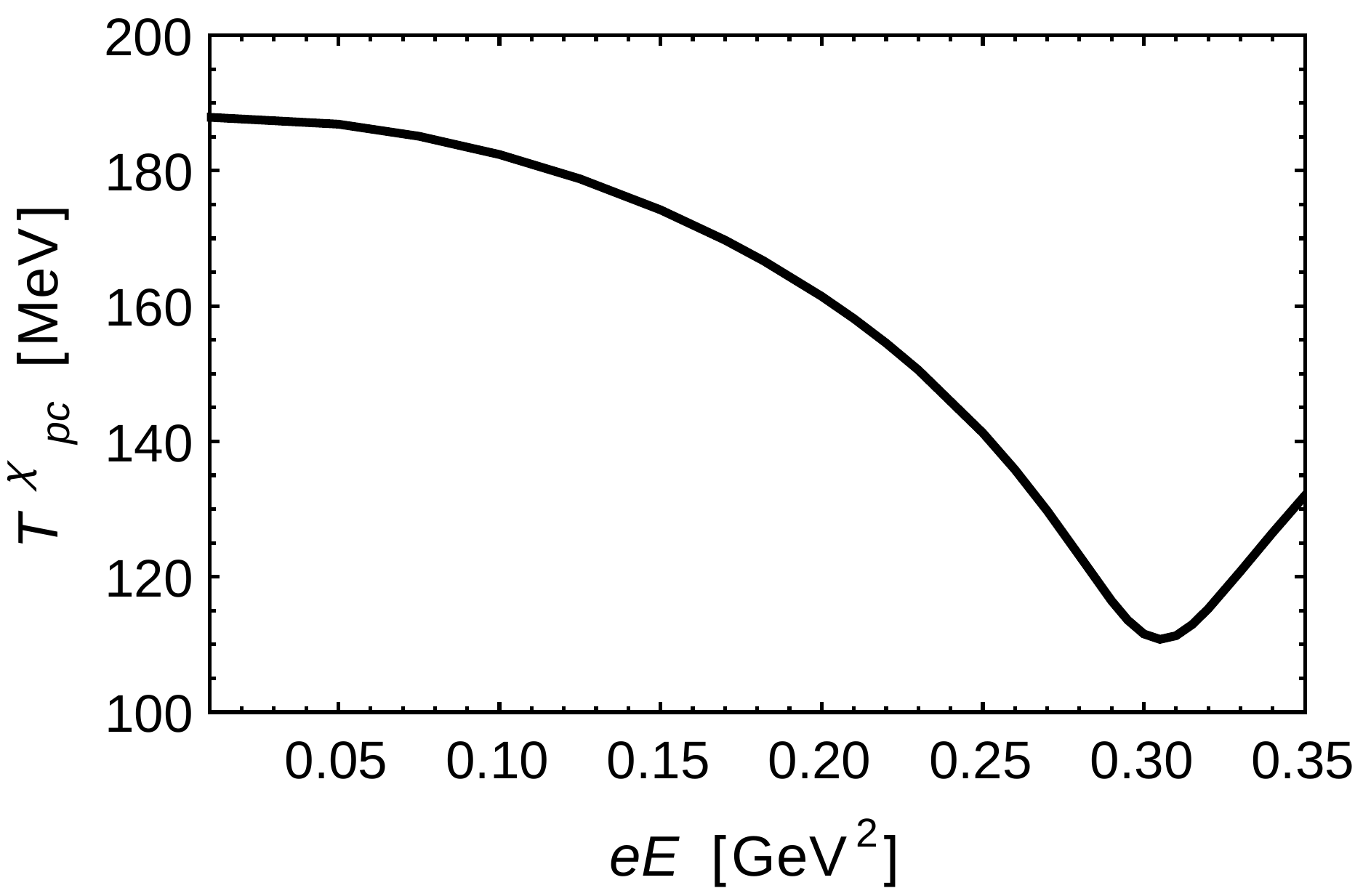}
\caption{Pseudocritical temperature for chiral symmetry restoration
  $T^{\chi}_{pc}$ as a function of $eE$ in the SU(2) NJL model. }
\label{fig3}
\end{figure}
\end{center}

In Fig.~\ref{fig1} we show the effective quark masses as a function of the temperature for different 
values of electric fields in the two-flavor NJL model. As expected, for
the NJL model at $eE=0.01$ GeV$^2$, as we increase the temperature, 
the effective quark masses decrease, 
as a signature of the chiral symmetry restoration .
If we increase the electric field
to $eE=0.10$ GeV$^2$, at low temperatures the effective quark masses slightly decrease as an effect of 
the restoration of the chiral symmetry guided by the electric field. This
effect was already explored in previous works in the literature at finite temperature~\cite{elec1,Cao} 
and at $T=0$ \cite{tatsumi,kleva}. As we increase the electric field it can be seen that the pseudocritial temperature has
 decreased  with the increase of the electric field. In Fig.\ref{fig2}, we show how $-\frac{dM}{dT}$ changes
with the increase of the electric field. The peak of each curve is interpreted as the pseudocritial temperature for chiral symmetry restoration $T_{pc}^{\chi}$ for the corresponding electric field. In Fig.\ref{fig3} 
we can see the pseudocritial temperature as a function of electric field for the two-flavor NJL model, showing the behavior previously predicted, that as we grow the electric field 
the pseudocritial temperature
decreases until we reach a critical value $eE = 0.31$ GeV$^2$. At this point, the pseudocritial temperature starts increasing as we increase the electric field. 

\begin{center}
\begin{figure}[!htb]
\includegraphics[width=7cm]{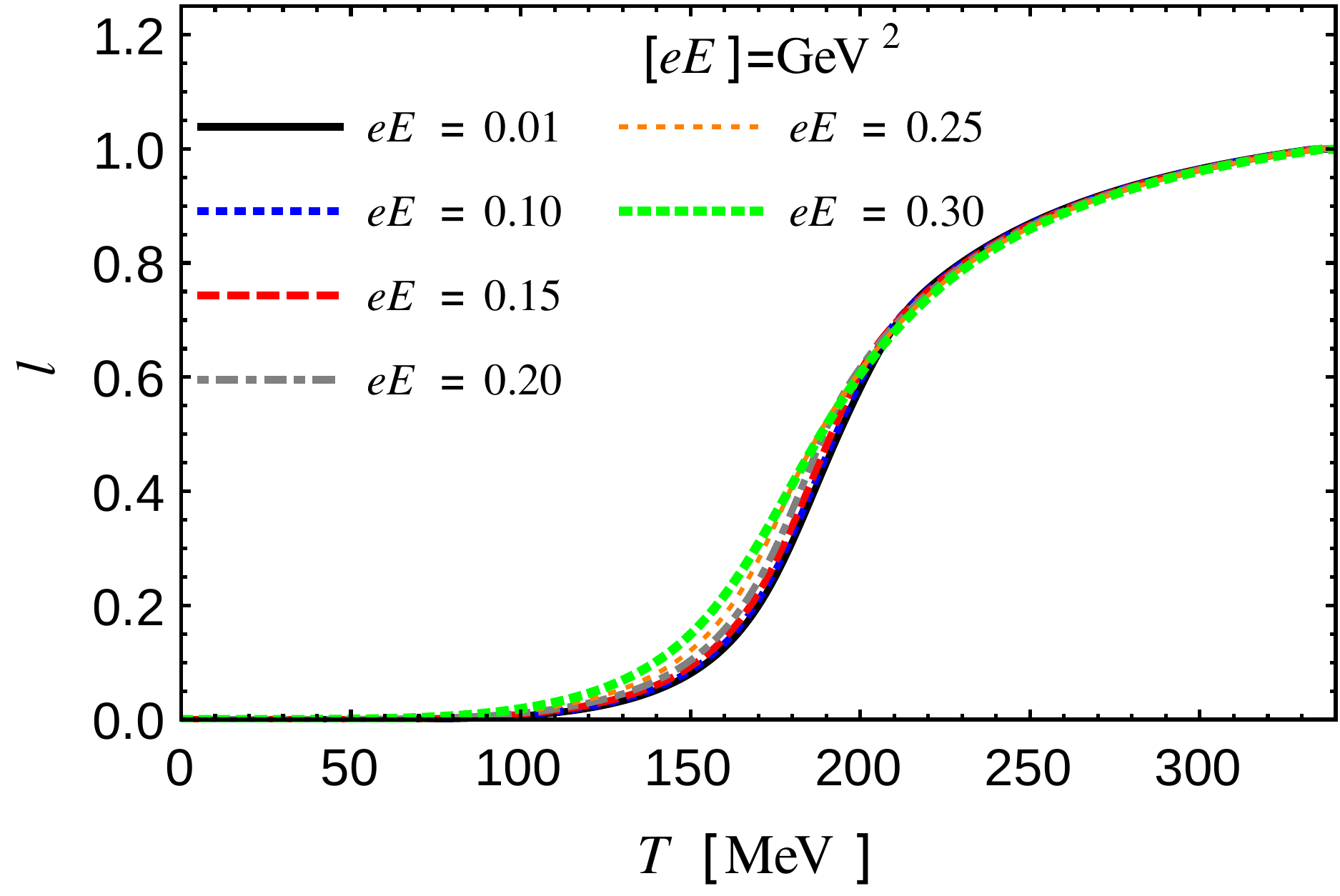}
\caption{Expectation value of the Polyakov loop for different values of $eE$ in the PNJL model. }
\label{fig4}
\end{figure}
\end{center}

 \noindent Next, we discuss our results for the SU(2) PNJL model. We show in Fig.~\ref{fig4} the
 expectation value of the Polyakov loop as function of the 
temperature for different values of 
electric fields. We can see that the effect of the strong electric fields
slightly changes the Polyakov loop expectation value in comparison with the effect on the
effective quark mass. These changes are more prominent at $T\sim 170$ MeV, where the
increase of the electric field tends to strengthen the deconfined phase. Also, we should mention that the
effect of the electric field in the Polyakov loop is given in an indirect way though the 
gap equation (\ref{gap2PNJL}) and Eq.(\ref{termgap}) (that are coupled), once the electric fields do
not couple directly with the Polyakov loop. 

\begin{center}
\begin{figure}[!htb]
\includegraphics[width=7cm]{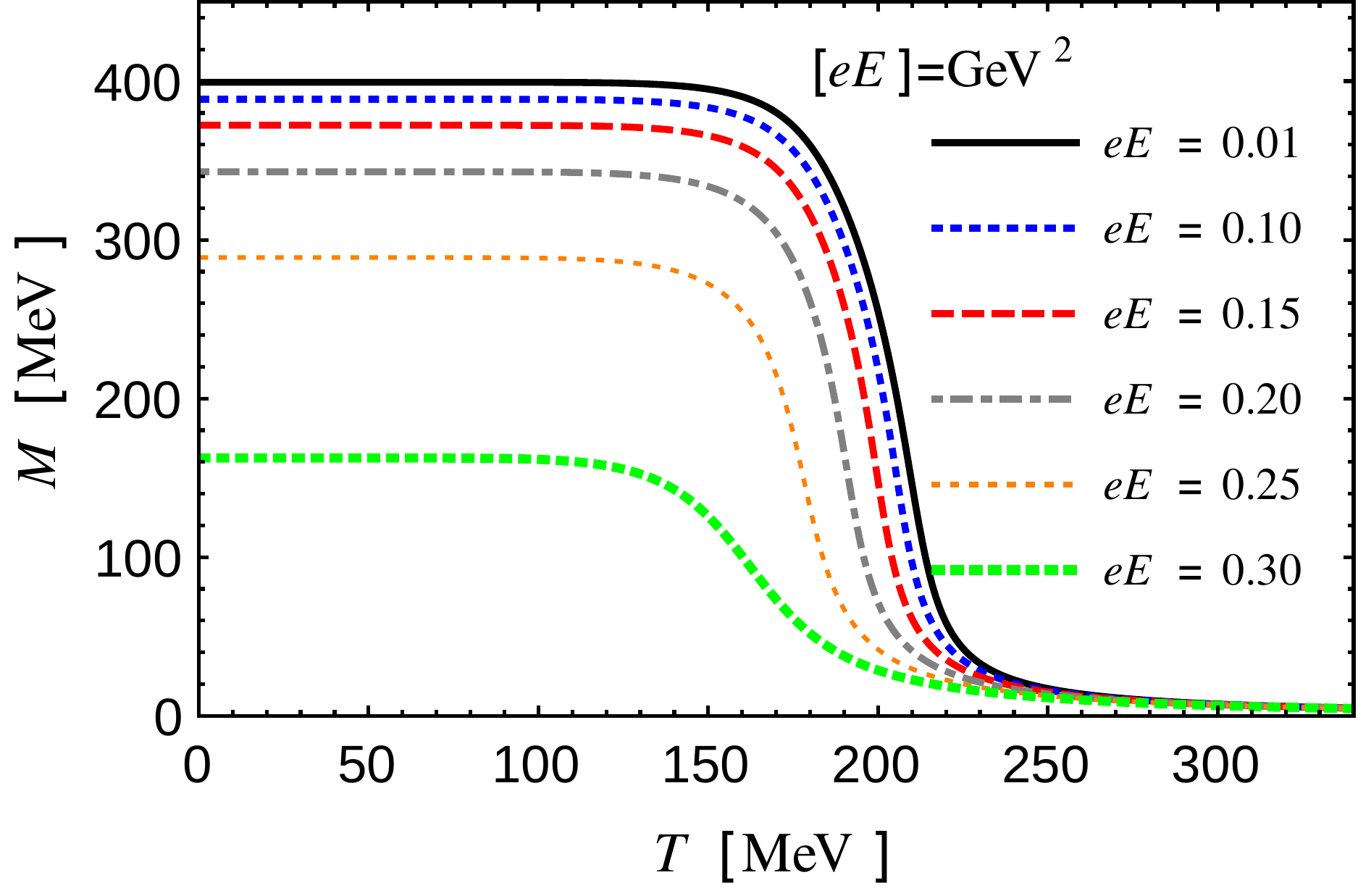}
\caption{Constituent quark mass as a function of $T$ for different values of $eE$ in the PNJL model. }
\label{fig5}
\end{figure}
\end{center}

\begin{center}
\begin{figure}[!htb]
\includegraphics[width=7cm]{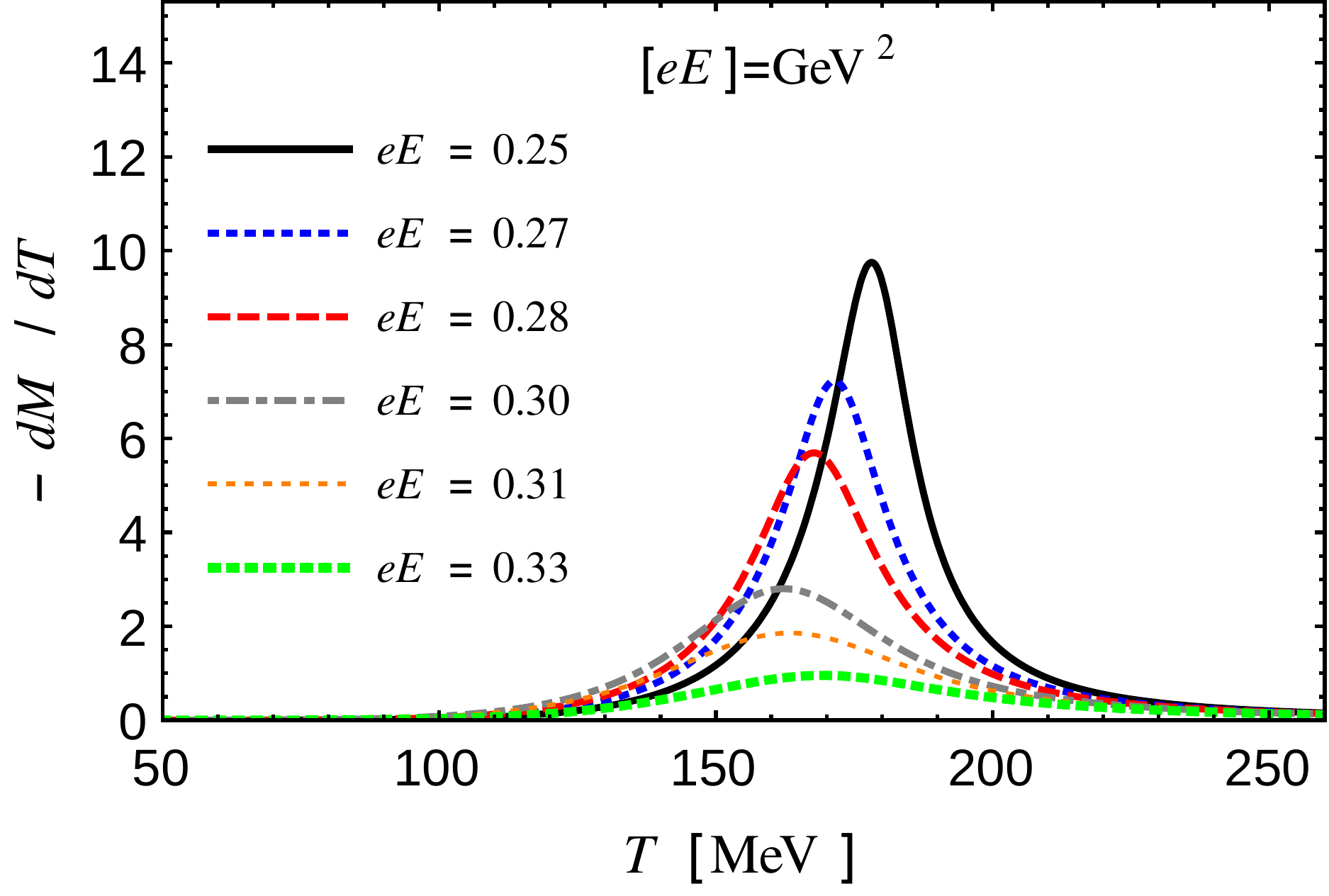}
\caption{$- \frac{d M}{dT}$ as functions of $T$ for different values of $eE$ in the PNJL model. }
\label{fig6}
\end{figure}
\end{center}

\begin{center}
\begin{figure}[!htb]
\includegraphics[width=7cm]{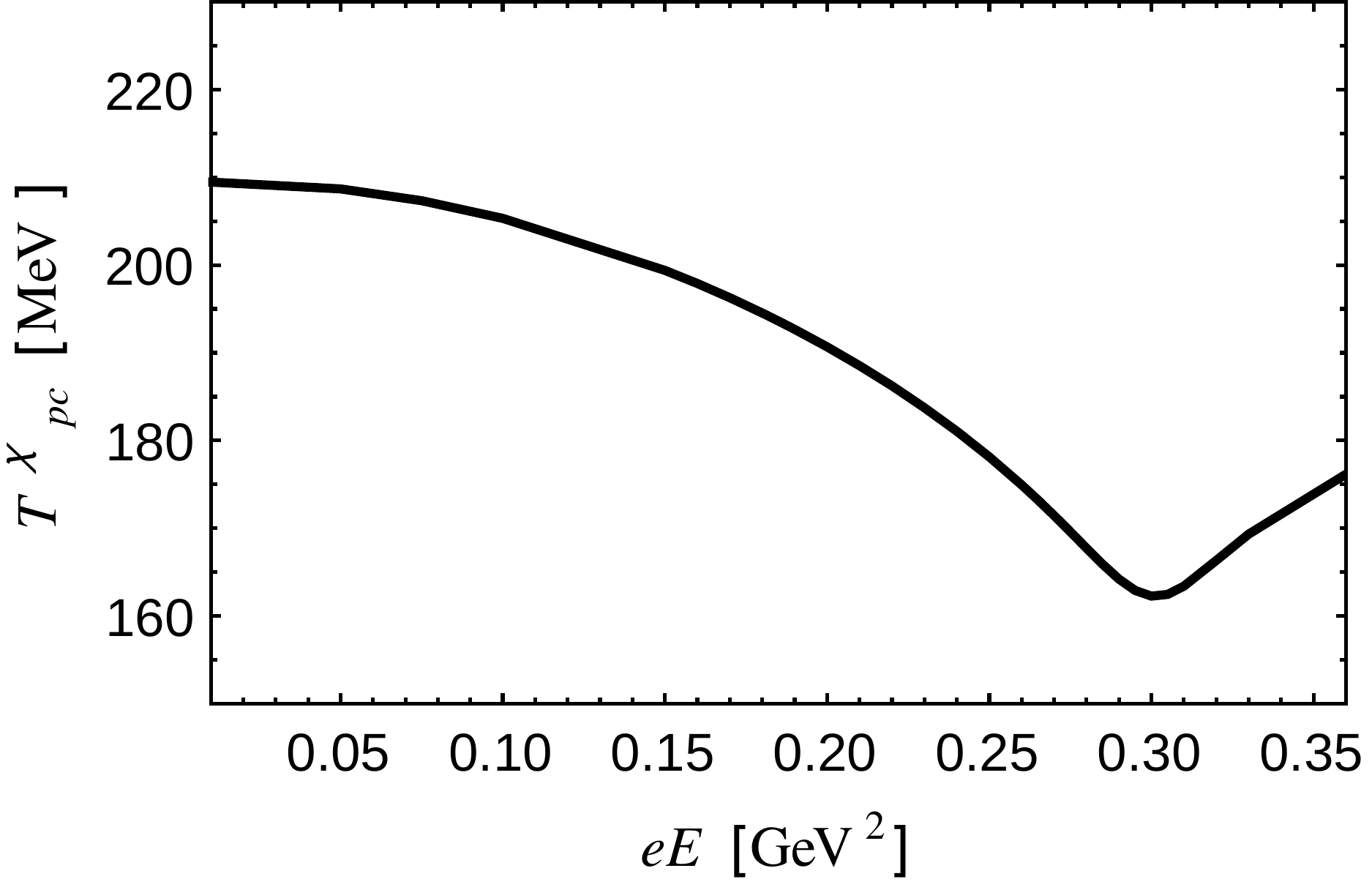}
\caption{Pseudocritical temperature $T_{pc}^\chi$ for chiral symmetry restoration of electrized 
quark matter as a function of the electric field $eE$ in the PNJL model.}
\label{fig7}
\end{figure}
\end{center}

In Fig.~\ref{fig5} we show the effective quark masses as function of the temperature for different values 
of electric fields in the two-flavor PNJL model, the behavior of $M$ as a function of $eE$ and $T$ 
is in qualitative agreement with that predicted in the NJL model. The quantity $-\frac{dM}{dT}$ for 
the two-flavor PNJL model is in Fig.~\ref{fig6} and 
in Fig.~\ref{fig7} we show the pseudocritial temperature for the chiral symmetry restoration as 
function of $T$, as we increase the electric field. The results obtained in PNJL model are in
quantitative agreement with NJL results, corroborating the idea that the present results
are model independent.

\begin{center}
\begin{figure}[!htb]
\includegraphics[width=7cm]{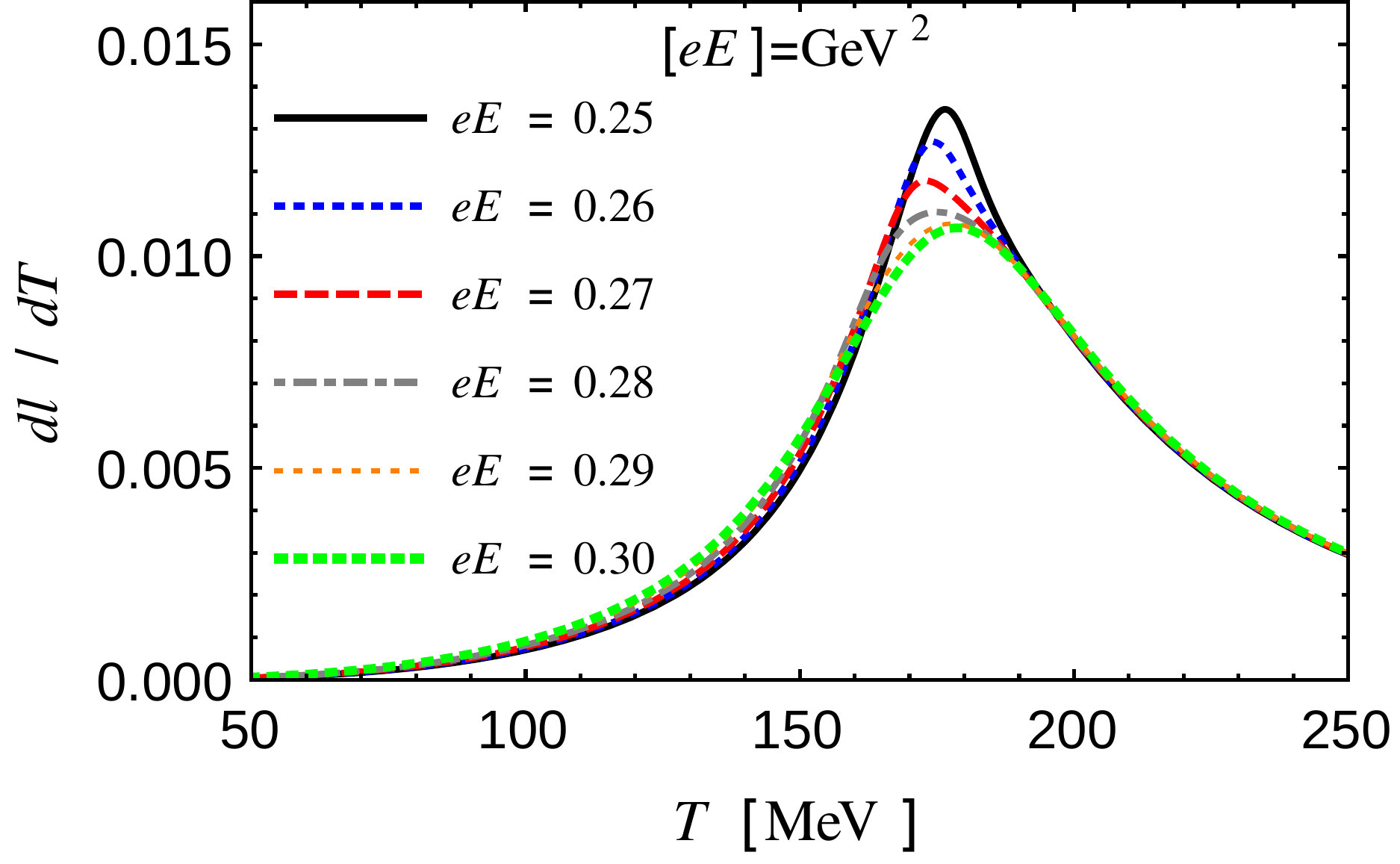}
\caption{$\frac{d l}{dT}$ as functions of $T$ for different values of $eE$ in PNJL model.}
\label{fig8}
\end{figure}
\end{center}

\begin{center}
\begin{figure}[!htb]
\includegraphics[width=7cm]{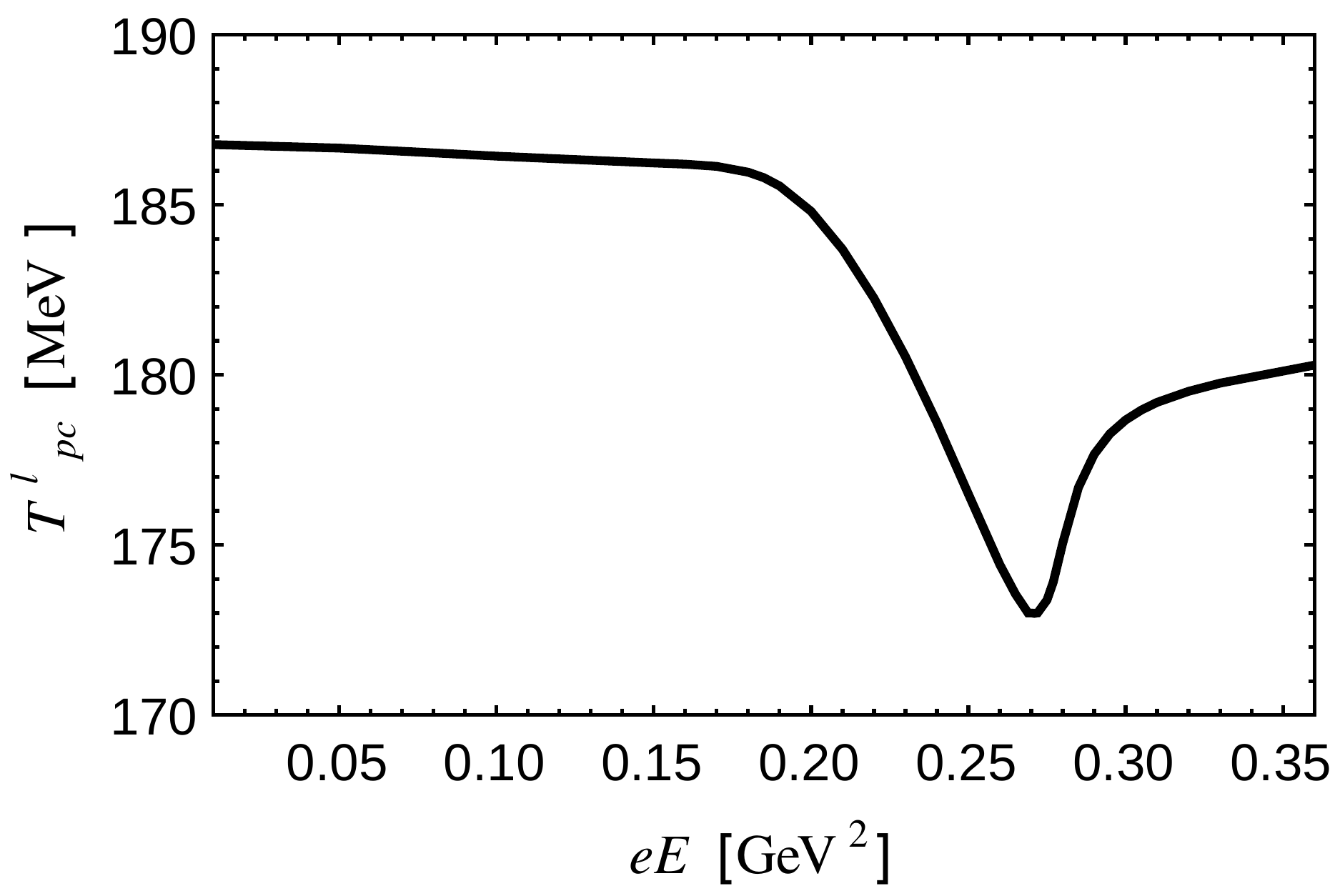}
\caption{The pseudocritial temperature $T_{pc}^\phi$ for the deconfinement transition
of electrized quark matter as a function of the electric field $eE$ in the PNJL model.}
\label{fig9}
\end{figure}
\end{center}

The quantity $\frac{dl}{dT}$ as a function of the temperature for different values of electric fields is shown in 
Fig.~\ref{fig8}; the  pseudocritial temperature for deconfinement $T_{pc}^l$ corresponds to the maximum of
each curve for different values of electric fields. In Fig.~\ref{fig9} we show $T_{pc}^l$ as a function of the electric field. 
The pseudocritial temperature for the deconfinement transition slightly decreases as we increase the electric field until $eE \sim 0.270$ GeV$^2$. At this point, 
the deconfinement temperature transition starts to increase in a similar way that we have predicted for the pseudocritial temperature for chiral symmetry restoration.

\begin{center}
\begin{figure}[!htb]
\includegraphics[width=7cm]{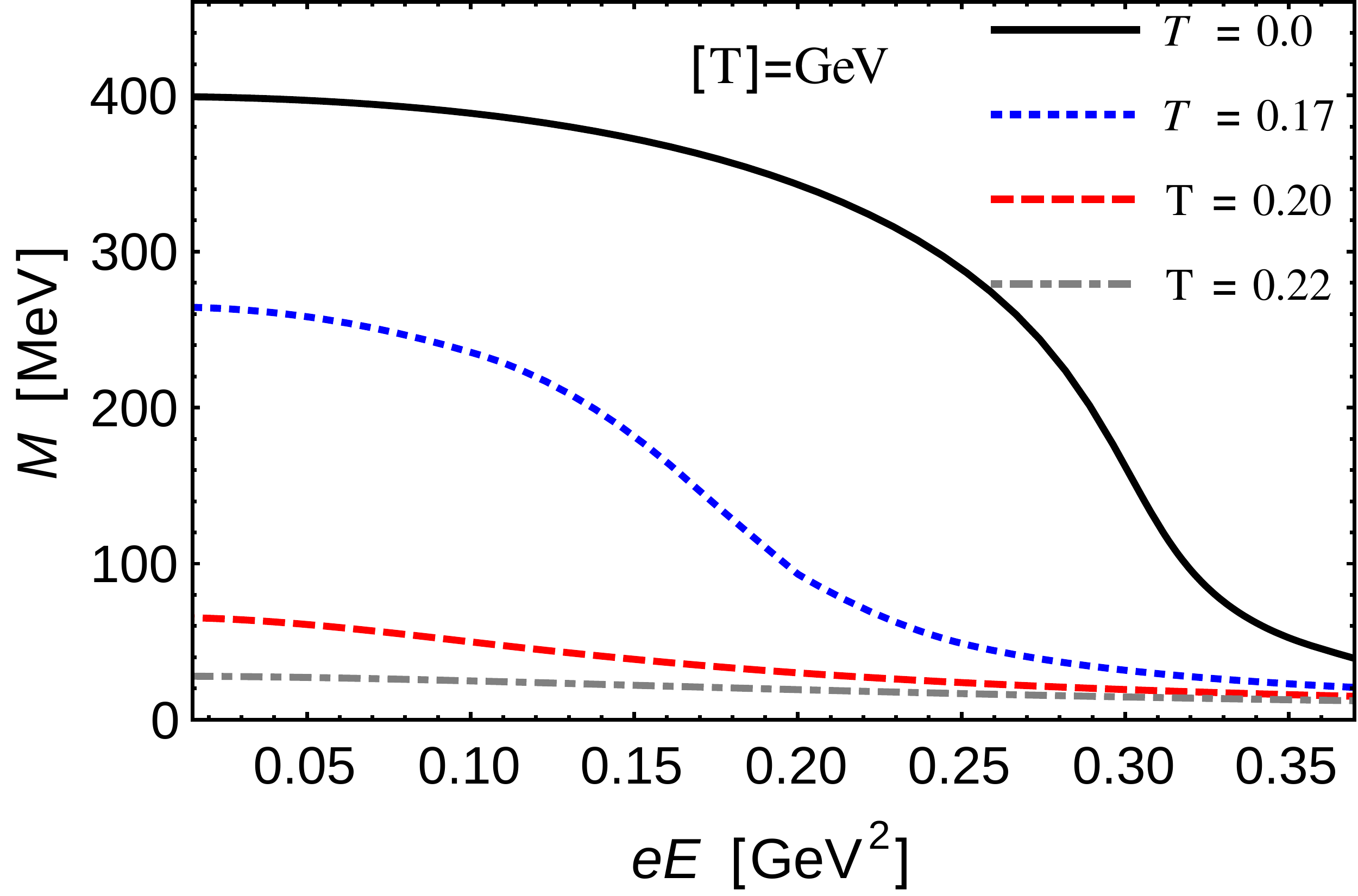}
\caption{Effective quark masses as a function of the electric field for fixed values of the temperature in the NJL SU(2) model.}
\label{fig10}
\end{figure}
\end{center}

  In Fig.~\ref{fig10} for the NJL SU(2) model the variation of the effective quark masses as a function of the electric field is shown for fixed temperatures 
: $T=0, 170, 200$, and $220$ MeV. 
At low values of the electric fields $eE\sim 0$, we can see the temperature effect on the partial restoration of the chiral symmetry.
 As we increase the magnitude of the electric field, the general aspect is the restoration of the chiral symmetry with the electric field.


\begin{center}
\begin{figure}[!htb]
\includegraphics[width=7cm]{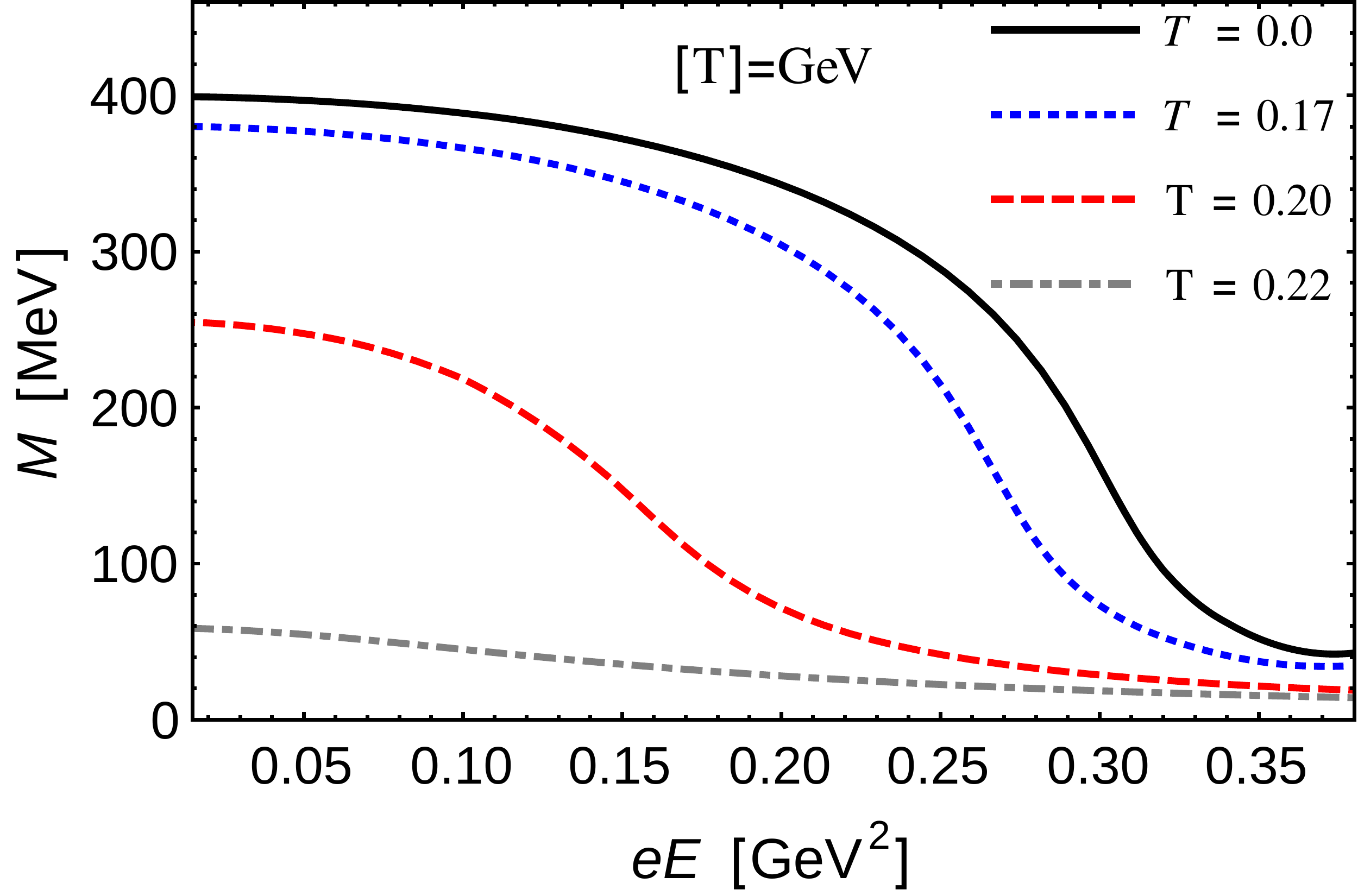}
\caption{Effective quark masses as a function of the electric field for fixed values of the temperature in the PNJL SU(2) model.}
\label{fig11}
\end{figure}
\end{center}

\begin{center}
\begin{figure}[!htb]
\includegraphics[width=7cm]{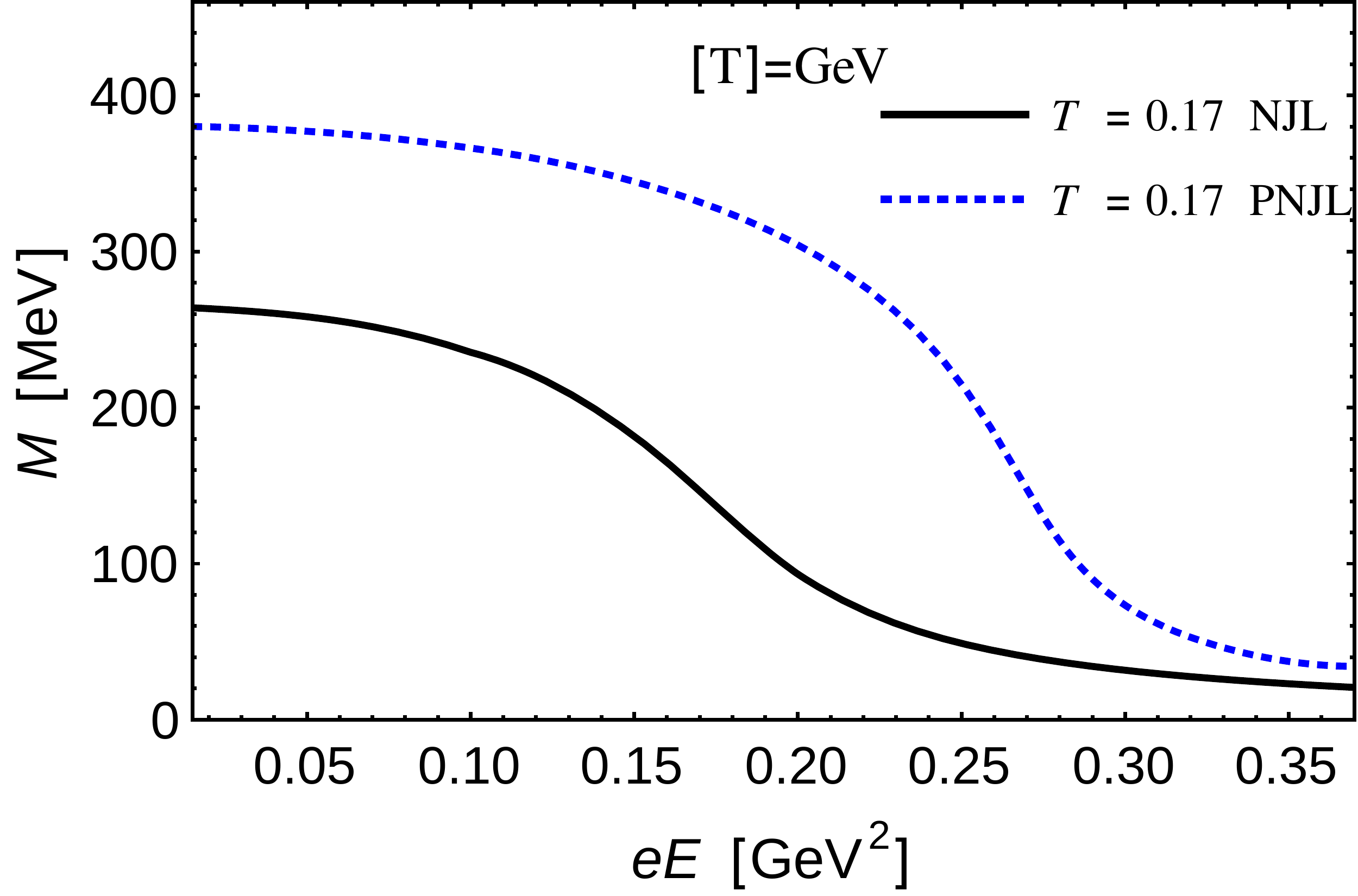}
\includegraphics[width=7cm]{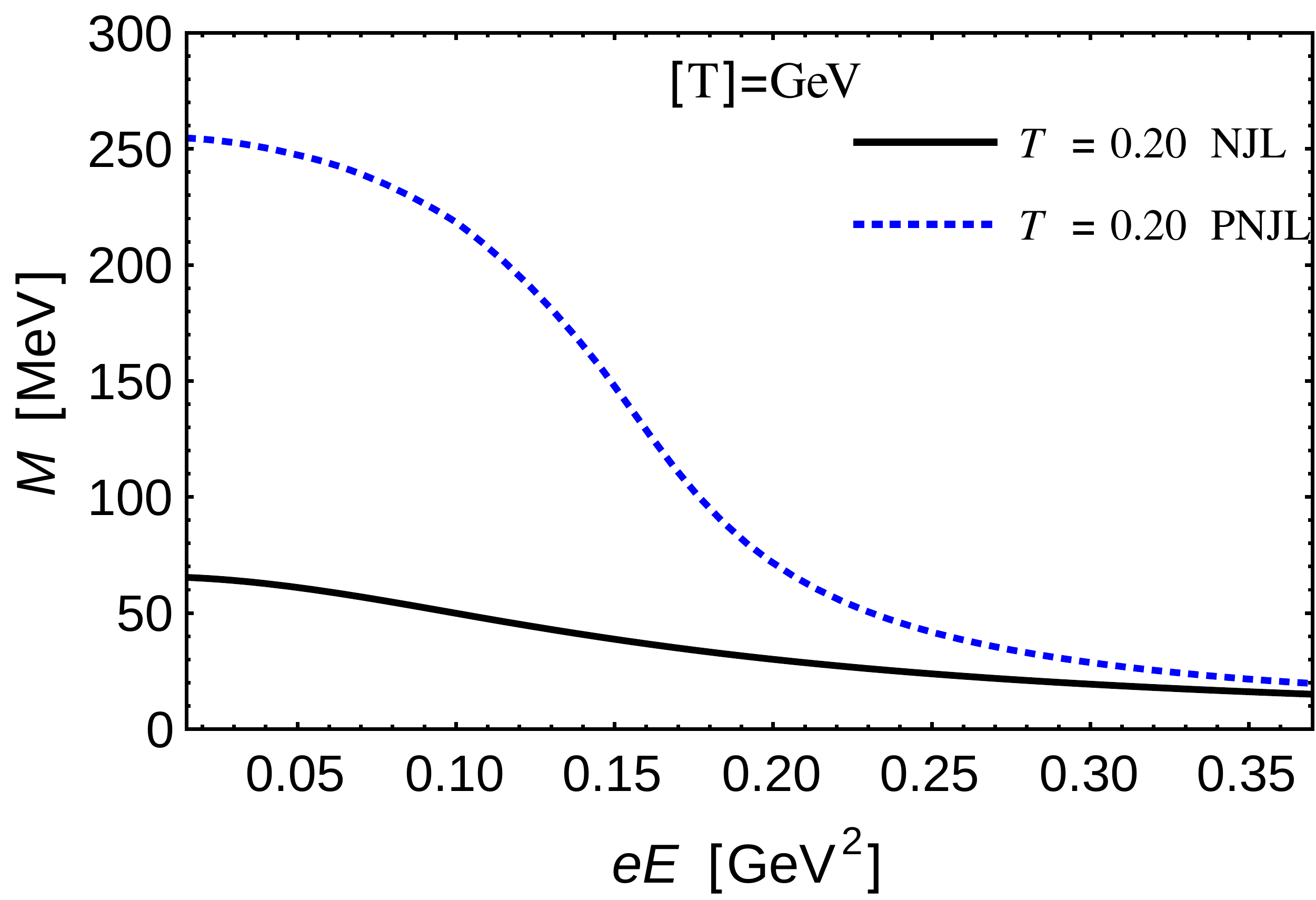}
\caption{Effective quark masses as a function of the electric field for fixed values of the temperature in the NJL SU(2) model compared with the PNJL SU(2) model 
for $T=170$ MeV (top panel) and 
$T=200$ MeV (lower panel).}
\label{fig12}
\end{figure}
\end{center}

  For the PNJL SU(2) results we have almost the same analysis, but the quantitative results are different as we can see in Fig.~\ref{fig11}. 
 In Fig.~\ref{fig12},  where we compare
 the two models at $T=170$ MeV and $T=220$ MeV, and we can see quantitative differences on the numerical results.  At $T=170$ MeV the
 NJL SU(2) model partially restores the chiral symmetry with a lower electric field than the PNJL SU(2) model. In the lower panel of Fig.~\ref{fig12} we have 
 the comparison of the models at $T=200$ MeV. We see that the PNJL SU(2) model has a much higher value of the effective quark mass at $eE\sim 0$ than the NJL SU(2). Also, 
 both models tends to partially restore the chiral symmetry to higher values of electric fields, following the previous analysis.
 At $T=0$ almost no difference is seen.
 All these results show quantitative differences between the two models and how
 the confinement can change the scenario of the restoration of the chiral symmetry.

%

 
It is interesting to point out the prediction of the differences of the effects of the electric fields on
$T_{pc}^l$ and $T_{pc}^{\chi}$ in the PNJL model. The electric fields tends to influence more easily the chiral transition than the deconfinement. 
The behavior of the chiral condensate and the Polyakov loop in an environment with constant magnetic field has been explored in~\cite{gatto1}.

%
\begin{center}
\begin{figure}[!htb]
\includegraphics[width=7cm]{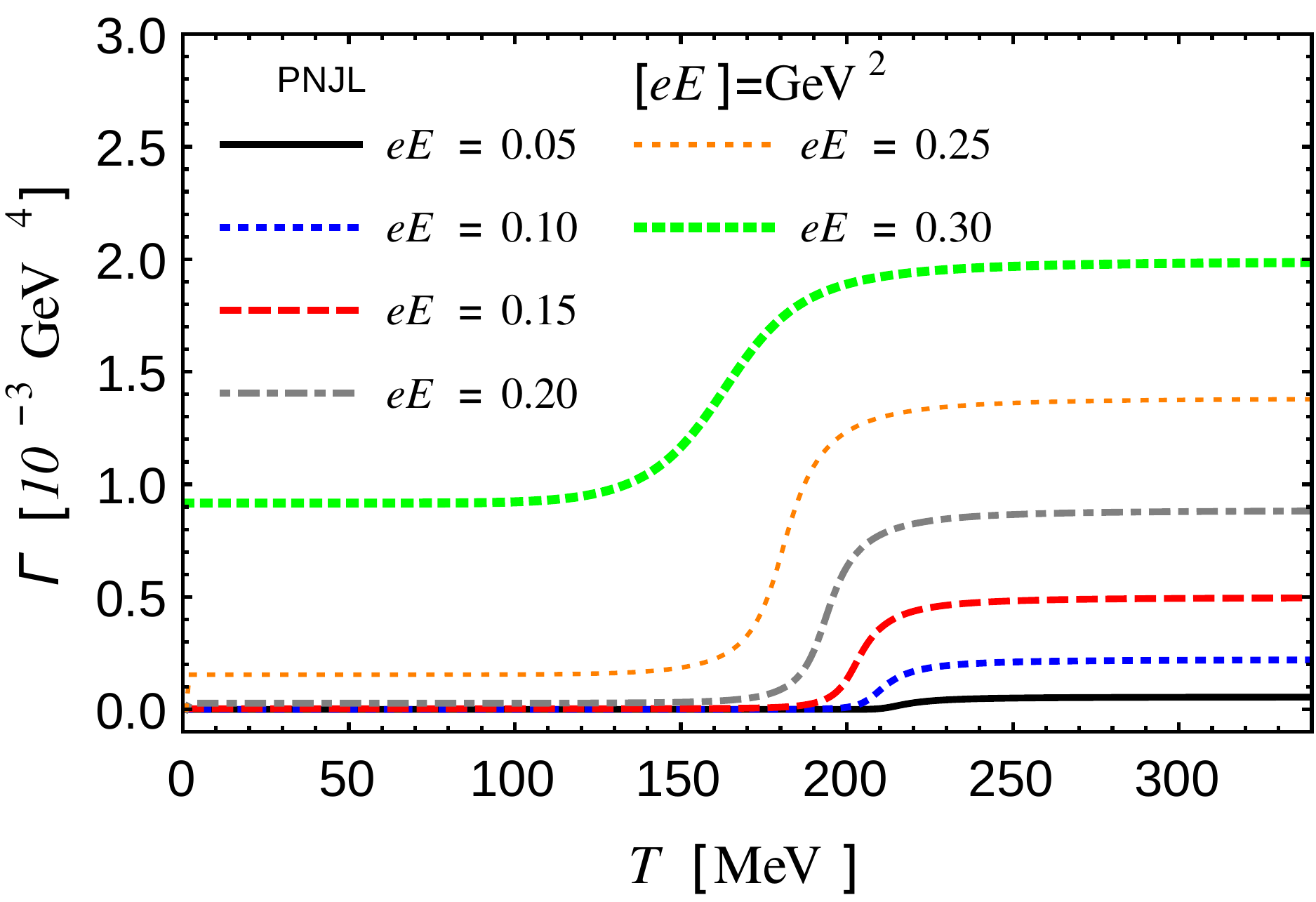}
\includegraphics[width=7cm]{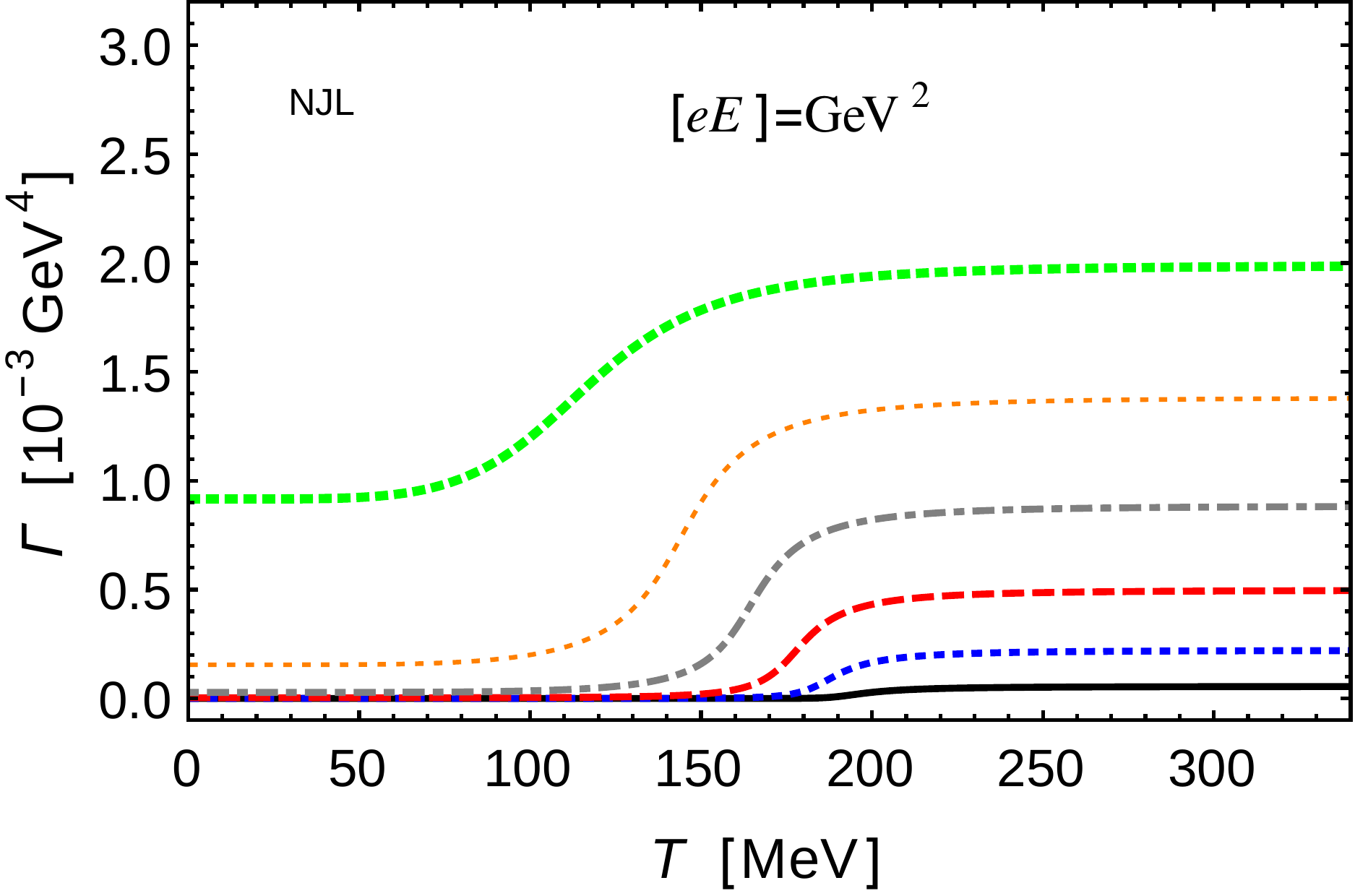}
\caption{Top: The pair production $\Gamma$ of electrized quark matter as a function of the temperature for different values of electric fields in the PNJL model. Bottom:
the same values of $eE$ for NJL model.}
\label{fig13}
\end{figure}
\end{center}


We see that the difference between the pseudocritial temperatures for chiral symmetry restoration at $eE=0.15$ GeV$^2$ and $eE=0.2$ GeV$^2$ is about $\Delta T_{pc}^{\chi}\sim 8.7$ MeV.
In the same manner, the difference in the deconfinement temperature is given by $\Delta T_{pc}^l \sim 1.43$ MeV.
As a conclusion, the quarks can be in a deconfined phase with chiral symmetry still not restored. This is a very interesting result, in the opposite 
direction of what happens when we have the quarkyonic phase~\cite{lowe,mclerran2}, that occurs for some values of $\mu\neq 0$ and $eE=eB=0$. At strong
enough electric fields, where both $T_{pc}^{\chi}$ and $T_{pc}^l$ increase as when we increase the 
electric field, the variation in the respective transition temperatures become larger and different. For example, 
if we take $eE=0.25$ GeV$^2$ and $eE=0.30$ GeV$^2$ we have, respectively, 
$\Delta T_{pc}^{\chi} \sim 15.88$ MeV  and $\Delta T_{pc}^l \sim 2.18$ MeV.
We should pay attention to the fact that we are working with magnitudes of the electrical fields that are
valid for the NJL and PNJL models $eE\sim \Lambda^2$. 

In Fig.\ref{fig13} we show the Schwinger pair production as
a function of the temperature at fixed electric fields
$eE = 0.05$ GeV$^2$, $eE = 0.10$ GeV$^2$, $eE = 0.15$ GeV$^2$, $eE = 0.20$ GeV$^2$, $eE = 0.25$ GeV$^2$ and $eE = 0.30$ GeV$^2$. In the region where we have chiral symmetry restoration we can see that the production rate grows, and after some value of temperature the Schwinger pair production stabilizes.
Another interesting aspect is that if you fix the electric field, in the NJL model the Schwinger pair production
stabilizes for larger values  and it happens 
in low temperatures if we compare with the PNJL model results. 
 
\section{Conclusions}
\label{conclusion}

In this work we have presented a study for strongly electrized quark matter within the the SU(2) PNJL and NJL 
models at finite temperatures in the mean field approximation. 
We have shown that the constituent quark masses decrease
as we grow both electric fields and temperatures, as a signature of the partial restoration of 
the chiral symmetry. 
In this scenario, as expected, the pseudocritial temperature for chiral symmetry restoration decreases 
as we grow the electric fields.
The deconfinement is guided by the expectation value of the Polyakov loop, and our results show that
the electric fields tend to anticipate the transition to the deconfined phase, and the effects due to 
the electric fields are more prominent in the chiral transition than the deconfinement one.
On the other hand, we show that for strong enough electric fields, both pseudocritial temperatures for chiral symmetry 
and deconfinement temperatures start to increase after a critical value of the electric field, a very interesting effect that has been identified for the first time in the literature.
This effect
propagates to all quantities, as the Schwinger pair production. 
For comparison, we also show the results in the NJL model, where the same qualitative results are obtained, 
revealing that the main characteristics of this type of theory are model independent. 
 Also, we observe that for some values of electric fields,
the quarks can be in a deconfined phase with the chiral symmetry still not restored, in the opposite direction of the observed quarkyonic phase for systems at finite baryonic density
and zero external fields.
We expect to use this type of phenomenology to extend the analysis for electromagnetic fields and more 
general purposes in the future, e.g., in systems that present a chiral imbalance of 
right-handed and left-handed quarks, finite baryonic densities and in the physics behind
the chiral magnetic effect~\cite{fukushima2,mssmu5,cme2}. Works in these directions are under way and we expect to report our results soon.

%
%
\appendix
\section{The trace of the Polyakov loop}
\label{app1}

The Polyakov loop can be expressed by a diagonal matrix $L=$diag($e^{i\varphi}$,$e^{i\varphi'}$,$e^{-i(\varphi+\varphi')}$). As 
discussed in~\cite{fuku0,andersen0}, 
the perturbative vacuum has $\varphi=\varphi'=0$, i.e., 
the $T=0$ limit, and the confining vacuum can be chosen to be $\varphi=\frac{2\pi}{3}$ and $\varphi'=0$. For 
simplicity, we can adopt $\varphi'=0$ from the beginning. 
The Polyakov loop with this assumption assumes $l=\frac{1}{N_c}\Tr_c L=\frac{1}{N_c}\Tr_c L^{\dagger}$. This is 
true for the limit $\mu=0$~\cite{fuku01}. 

The simplification adopted here implies $l=\frac{1}{3}(1+2\cos \varphi)$. It is useful to invert the last 
relation for future evaluations 

\begin{equation}
 \varphi=\cos^{-1}\left(\frac{3l-1}{2}\right).\label{phitrace}
\end{equation}

In this work we will implement the $\mu=0$ case (where we have $\bar{l}=l$) and $\Tr_c$, i. e., the trace
over the color space is performed following the steps given in
Eqs.(\ref{intrep}) and (\ref{repl}).  First we decompose the trace as

\begin{eqnarray}
\Tr_c[(L)^n+(L^{\dagger})^n]=\Tr_c (L)^n+\Tr_c (L^{\dagger})^n \label{trace} \, .
\end{eqnarray}

To evaluate the trace of $L^n$ and $(L^{\dagger})^n$ we should note that the Ansatz is already in the
Jordan form {\cite{horn}}; then we can use the following result:

\begin{eqnarray}
 \Tr A^n = \sum_{i}\lambda_i^n,
\end{eqnarray}

\noindent where the $\lambda_i$ are the eigenvalues of the matrix $A$.
 Applying this result directly to Eq.(\ref{trace}), we obtain

\begin{eqnarray}
\Tr_c (L)^n+\Tr_c (L^{\dagger})^n &&=2(1+e^{i\varphi n}+e^{-i\varphi n}),\nonumber\\
&&=2(1+2\cos n \varphi).
\end{eqnarray}

Using now Eq.(\ref{phitrace}) in the last equation, we can reach

\begin{eqnarray}
&&\Tr_c (L)^n+\Tr_c (L^{\dagger})^n = 2\nonumber\\
&&\times\left\{1+2\cos\left[n \cos^{-1}\left(\frac{3l-1}{2}\right) \right] \right\}.\nonumber\\
\end{eqnarray}

\section*{Acknowledgments}

This work is partially supported by Conselho Nacional de Desenvolvimento Cient\'{\i}fico 
e Tecnol\'ogico - CNPq, Grants. No. 304758/2017-5 (R.L.S.F), and No. 6484/2016-1 (S.S.A); as part of the project Instituto Nacional de Ci\^encia e Tecnologia - F\'{\i}sica Nuclear
e Aplica\c c\~oes (INCT-FNA) Grant. No. 464898/2014-5 (S.S.A.); and Funda\c{c}\~ao de Amparo \`a Pesquisa do Estado do  
Rio Grande do Sul (FAPERGS), Grant No. 19/2551-0000690-0 (R.L.S.F.); and Coordena\c c\~ao de Aperfei\c{c}oamento de Pessoal  de N\'ivel Superior (CAPES) (W.R.T.), Brasil (CAPES), Finance Code 001.


\vfill

\begin{thebibliography}{99}

\bibitem{skokov}A. Bzdak and V. Skokov, Phys.
Lett. B {\bf 710}, 171 (2012).

\bibitem{deng} W. T. Deng and X. G. Huang, Phys. Rev. C {\bf 85}, 044907 (2012).

\bibitem{zhang} J. Bloczynski, X. G. Huang, X. Zhang, and J. Liao, Phys. Lett. B {\bf 718}, 1529
(2013).

\bibitem{deng2} W. T. Deng and X. G. Huang, Phys. Lett. B {\bf 742}, 296 (2015).

\bibitem{fukushima2}  K. Fukushima, D. E. Kharzeev and H. J. Warringa, Phys. Rev. D {\bf 78}, 074033, (2008); D. E. Kharzeev and
H. J. Warringa, Phys. Rev. D {\bf 80}, 034028 (2009); D. E.
Kharzeev, Nucl. Phys. {\bf A830}, 543c (2009).


\bibitem{duncan}R. Duncan and C. Thompson, Astron. J., {\bf 392}, L9 (1992).

\bibitem{kouve} C. Kouveliotou {\it et al.}, Nature (London) {\bf 393}, 235 (1998).

\bibitem{zhang2} J. Bloczynski, X. G. Huang, X. Zhang, and J. Liao, Nucl. Phys. {\bf A939}, 85 (2015).

\bibitem{hirono} Y. Hirono, M. Hongo, and T. Hirano, Phys. Rev. C {\bf 90}, 021903(R) (2014).

\bibitem{Voronyuk} V. Voronyuk, V. D. Toneev, S. A. Voloshin, and W. Cassing, Phys. Rev. C {\bf 90}, 064903 (2014).

\bibitem{cheng2} Yi-Lin Cheng, Song Zhang, Yu-Gang Ma, Jin-Hui Chen, and Chen Zhong, Phys. Rev. C {\bf 99}, 054906 (2019).

\bibitem{sheng} Hui Li, Xin-li Sheng, and Qun Wang, Phys. Rev. C {\bf 94}, 044903 (2016).

\bibitem{review} J. Zhao and F. Wang,
Prog. Part. Nucl. Phys. {\bf 107}, 200 (2019).

\bibitem{xu1} X. G. Huang and J. Liao, Phys. Rev. Lett. {\bf 110}, 232302 (2013). 

\bibitem{xu2} Y. Jiang, X. G. Huang, and J. Liao, Phys. Rev. D {\bf 91}, 045001 (2015).

\bibitem{shipu} Shi Pu, Shang-Yu Wu, and Di-Lun Yang
Phys. Rev. D {\bf 91}, 025011 (2015).

\bibitem{shipu2} Shi Pu, Shang-Yu Wu, and Di-Lun Yang
Phys. Rev. D {\bf 89}, 085024 (2014).

\bibitem{review2} X. G. Huang, Rep. Prog. Phys. {\bf 79}, 076302 (2016).

\bibitem{shovkovy}V. A. Miransky and I. A. Shovkovy, Phys. Rep. {\bf 576}, 1 (2015); D. Ebert, K. G. Klimenko, M. A. Vdovichenko, and A. S. Vshivtsev, Phys. Rev. D {\bf 61}, 025005 (1999); 
K. G. Klimenko, Z. Phys. C {\bf 54}, 323 (1992); arXiv:hep-ph/9809218.

\bibitem {avancini2}D. P. Menezes, M. Benghi Pinto, S. S. Avancini, and C. Provid\^encia, Phys. Rev. C {\bf 80}, 065805 (2009).


\bibitem{nosso1} D. P. Menezes, M. Benghi Pinto, S. S. Avancini, A. P\'erez Mart\'inez, and C. Provid\^encia, Phys. Rev. C {\bf 79}, 035807 (2009).

%
\bibitem{Farias:2014eca} 
  R.~L.~S.~Farias, K.~P.~Gomes, G.~I.~Krein, and M.~B.~Pinto,
  Phys.\ Rev.\ C {\bf 90}, 025203 (2014).
%
\bibitem{Farias:2016gmy} 
  R.~L.~S.~Farias, V.~S.~Timoteo, S.~S.~Avancini, M.~B.~Pinto, and G.~Krein,
  Eur.\ Phys.\ J.\ A {\bf 53}, 101 (2017).
%
\bibitem{Avancini:2016fgq} 
  S.~S.~Avancini, R.~L.~S.~Farias, M.~Benghi Pinto, W.~R.~Tavares, and V.~S.~Tim\'oteo,
  Phys.\ Lett.\ B {\bf 767}, 247 (2017).
%
\bibitem{Avancini:2018svs} 
  S.~S.~Avancini, R.~L.~S.~Farias, and W.~R.~Tavares,
  Phys.\ Rev.\ D {\bf 99}, 056009 (2019).
%
\bibitem{Ayala_LSM1} 
A. Ayala, M. Loewe,  A. Z. Mizher, and R. Zamora,
Phys. Rev.D {\bf 90}, 036001  (2014).
%
\bibitem{Ayala_LSM2} 
A.~Ayala, M.~Loewe, and R.~Zamora,
Phys.\ Rev.\ D {\bf 91}, 016002 (2015).
%
\bibitem{bali} G. Bali, F. Bruckmann, G. Endr\"{o}di, Z. Fodor, S. Katz, {\it et al.}, I. High Energy Phys. 02 (2012) 044.

\bibitem{endrodiIMC} G. S. Bali, F. Bruckmann, G. Endr\"{o}di, Z. Fodor, S. D. Katz, and A. Sch\"{a}fer, Phys. Rev. D {\bf 86}, 071502(R) (2012).

\bibitem{fukushima} K. Fukushima and T. Hatsuda, Rep. Prog. Phys. {\bf 74}, 014001 (2011).

\bibitem{gatto1} K. Fukushima, M. Ruggieri and R. Gatto, Phys. Rev. D {\bf 81}, 114031 (2010).

\bibitem{gatto2} R. Gatto and M. Ruggieri, Phys.Rev. D {\bf82}, 054027 (2010).

\bibitem{gatto3} R. Gatto and M. Ruggieri, Phys.Rev. D {\bf 83}, 034016 (2011).

\bibitem{kashiwa} K. Kashiwa, Phys.Rev. D {\bf 83}, 117901 (2011).

\bibitem{endrodi} G. Endr\"{o}di and G.J. Marko, J. High Energy Phys. 08 (2019) 036. 

\bibitem{mclerran} L.D Mclerran and B. Svetitsky, Phys. Rev. D {\bf 24}, 450 (1981).

\bibitem{cheng} M. Cheng  {\it et al.}, Phys. Rev. D {\bf 77}, 014511 (2008).

\bibitem{ratti01} C. Ratti, S. R{\H o}ssner, M. Thaler, and W. Weise, Eur. Phys. J. C {\bf 49}, 213 (2007).

\bibitem{ratti0} C. Ratti, M. A. Thaler, and W. Weise, Phys. Rev. D {\bf 73}, 014019 (2006).

\bibitem{fukushima03} K. Fukushima and C. Sasaki, Prog. Part. Nucl. Phys. {\bf 72}, 99 (2013).

\bibitem{ratti} H. Hansen, W.M. Alberico, A. Beraudo, A. Molinari, M. Nardi, and C. Ratti, Phys. Rev. D {\bf 75} 065004 (2007).

\bibitem{andersen0} J. O. Andersen, W. R. Naylor, and A. Tranberg, Rev. Mod. Phys. {\bf 88}, 025001 (2016).

\bibitem{nambu}Y. Nambu and G. Jona-Lasinio, Phys. Rev. {\bf 122}, 345 (1961); {\bf 124}, 246 (1961).

\bibitem{elec1} W.R. Tavares and S.S. Avancini, Phys. Rev. D {\bf 97}, 094001 (2018). 

\bibitem{Cao} G. Cao and X. G. Huang, Phys. Rev. D {\bf 93}, 016007 (2016). 

\bibitem{ruggieri1} M. Ruggieri, Z.Y. Lu, and G.X.Peng, Phys. Rev. D {\bf 94}, 116003 (2016).

\bibitem{ruggieri2} M. Ruggieri and G.X. Peng, Phys. Rev. D {\bf 93}, 094021 (2016).

\bibitem{Schwinger} J. S. Schwinger, Phys. Rev. {\bf 82}, 664 (1951).

\bibitem{heisenberg} W. Heisenberg and H. Euler, Z. Phys. {\bf 98}, 714 (1936).

\bibitem{fukuansatz} K. Fukushima, Phys. Rev. D {\bf 77}, 114028 (2008).

\bibitem{lowe} J. P. Carlomagno and M. Loewe, Phys. Rev.D {\bf 95} 036003 (2017).

\bibitem{kleva} S. P. Klevansky, Rev. Mod. Phys. {\bf 64}, 649 (1992); S. P. Klevansky and R. H. Lemmer, Phys. Rev. D {\bf 39}, 3478 (1989).

\bibitem{buballa} M. Buballa, Phys. Rep. {\bf 407}, 205 (2005).

\bibitem{ayala} A. Ayala, C. A Dominguez, L. A. Hernandez, M. Loewe, A. Raya, J. C. Rojas, and C. Villavicencio, Phys. Rev. D {\bf 94}, 054019 (2016). 

\bibitem{paw} B.-J. Schaefer, J. M. Pawlowski, and J. Wambach, Phys. Rev. D {\bf 76}, 074023 (2007).

\bibitem{tatsumi} H. Suganuma and T. Tatsumi, Prog. Theor. Phys. {\bf 90}, 379 (1993).

\bibitem{mclerran2}  L. McLerran, K. Redlich, and C. Sasaki, Nucl. Phys. {\bf A824}, 86 (2009).

\bibitem{mssmu5} R.L.S. Farias, D.C. Duarte, G. Krein, and R.O. Ramos, Phys. Rev. D {\bf 94}, 074011 (2016).

\bibitem{cme2} D. E. Kharzeev, L. D. McLerran, and H. J. Warringa, Nucl.
Phys. {\bf A803}, 227 (2008).

\bibitem{fuku0} K. Fukushima, Phys. Lett. B{\bf 591}, 277 (2004).

\bibitem{fuku01}H. Abuki and K. Fukushima, Phys. Lett. B {\bf 676}, 57 (2009).

\bibitem{horn} A.R. Horn, and C.R. Johnson, \textit{Matrix Analysis}(Cambridge University Press, New York, 2013), 2nd ed.


\end{thebibliography}
\end{document}